# NMR Studies of Adsorption and Diffusion in Porous Carbonaceous Materials

Alexander C. Forse,[1] Céline Merlet,[2,3] Clare P. Grey,[1] John M. Griffin,[4*]

1. Department of Chemistry, University of Cambridge, Lensfield Road, Cambridge, CB2 1EW, UK
2. CIRIMAT, Université de Toulouse, CNRS, Université Toulouse 3 - Paul Sabatier, 118 Route de Narbonne, 31062 Toulouse cedex 9, France
3. Réseau sur le Stockage Électrochimique de l'Énergie (RS2E), Fédération de Recherche CNRS 3459, HUB de l'Énergie, Rue Baudelocque, 80039 Amiens, France
4. Department of Chemistry, Lancaster University, Lancaster, LA1 4YB, UK

**Abstract**

Porous carbonaceous materials have many important industrial applications including energy storage, water purification, and adsorption of volatile organic compounds. Most of their applications rely upon the adsorption of molecules or ions within the interior pore volume of the carbon particles. Understanding the behaviour and properties of adsorbate species on the molecular level is therefore key for optimising porous carbon materials, but this is very challenging owing to the complexity of the disordered carbon structure and the presence of multiple phases in the system. In recent years, NMR spectroscopy has emerged as one of the few experimental techniques that can resolve adsorbed species from those outside the pore network. Adsorbed, or "in-pore" species give rise to resonances that appear at lower chemical shifts compared to their free (or "ex-pore") counterparts. This shielding effect arises primarily due to ring currents in the carbon structure in the presence of a magnetic field, such that the observed chemical shift differences upon adsorption are nucleus-independent to a first approximation. Theoretical modelling has played an important role in rationalising and explaining these experimental observations. Together, experiments and simulations have enabled a large amount of information to be gained on the adsorption and diffusion of adsorbed species, as well as on the structural and magnetic properties of the porous carbon adsorbent. Here, we review the methodological developments and applications of NMR spectroscopy and related modelling in this field, and provide perspectives on possible future applications and research directions.

## 1. Introduction

Porous carbons are an important class of materials which have a wide range of industrial and technological applications including energy storage, filtration, gas storage and catalysis.[1-5] Most applications exploit the ability of porous carbons to adsorb molecules and ions on their internal surfaces, which are accessed via the pore network. In principle, carbon pore structures and topologies can be empirically optimised for a particular application, with pore sizes ranging from the microporous (diameters < 2 nm) to mesoporous (diameters 2 - 50 nm) to macroporous (diameters > 50 nm) regimes (or a distribution of sizes), and surface areas ranging from several hundred to more than 2000 $m^2\ g^{-1}$, depending on the synthesis procedure and precursor materials used. However, to improve the properties of porous carbons, it is necessary to have a detailed understanding of the behaviour of the adsorbate species within the pore network and how they interact with the carbon surface. This is a very challenging task owing to the complexity of most porous carbon-adsorbate systems. Many porous carbon structures are amorphous or lack long-range order beyond a few carbon-carbon bond lengths, making it difficult to extract structural information. Adsorbates are also challenging to study, as any technique must be capable of distinguishing adsorbed species from any non-adsorbed species that are present, as well as from the large amount of solid carbon in the system. Furthermore, most adsorbates remain highly dynamic within the pore network, making it difficult to characterise or define any ordering that is present, and any information that is obtained must be contextualised within a particular timescale.

Over the last 10 years, NMR spectroscopy has re-emerged as a highly sensitive technique for probing the behaviour of adsorbates in porous carbons. Although its utility for this purpose was first demonstrated over 25 years ago, the recent resurgence of interest in porous carbon applications, particularly for energy storage applications, has reinvigorated the development of NMR for this purpose. Although NMR is by no means the only technique capable of studying these systems, it has a number of advantages that enable it to provide a unique perspective, providing quantitative local information about adsorbate species. In particular, the fact that NMR has no requirement for long-range order and is applicable to both solid and liquid samples, means it is well-suited for studying liquid or gaseous species adsorbed on porous carbons which may be highly mobile. In addition the fact that NMR is intrinsically element specific means that information can be obtained for one species of interest without complication from the other components of the system. Furthermore, the sensitivity of NMR to dynamics over a wide range of timescales means that various aspects of the dynamic behaviour of adsorbates can be probed.

This review summarises recent developments in NMR spectroscopy for the study of adsorbates in porous carbons in terms of both the experimental advances that have been made, and the new understanding that has been gained. The aim is to give an overview of the current place of NMR spectroscopy within porous carbons research, and also to highlight areas for further development or study.

## 2. Structures and properties of carbonaceous materials

### *2.1 Structural and chemical variability in carbonaceous materials*

A large variety of carbon allotropes can be synthesized or found in nature. These allotropes have different structural, mechanical, and electronic properties, which can sometimes be tuned, making carbon materials suitable for a large number of applications. Figure 1 gathers a range of synthetic carbon materials with a range of different characteristics.[6] It is important to note at the outset that activated carbons are currently the most important porous carbon materials in industry, with these carbons produced on a large scale, and so these materials are the main focus of this review.

Carbon onions are nonporous nanoparticles while carbon nanotubes can be porous or nonporous depending on their diameter. Carbons onions and nanotubes offer a high conductivity and relatively large surface areas, around 500-600 $m^2 g^{-1}$ and 1000 $m^2 g^{-1}$ respectively. The highly accessible surfaces of these materials make them promising for applications where fast adsorption/desorption of molecules is important.

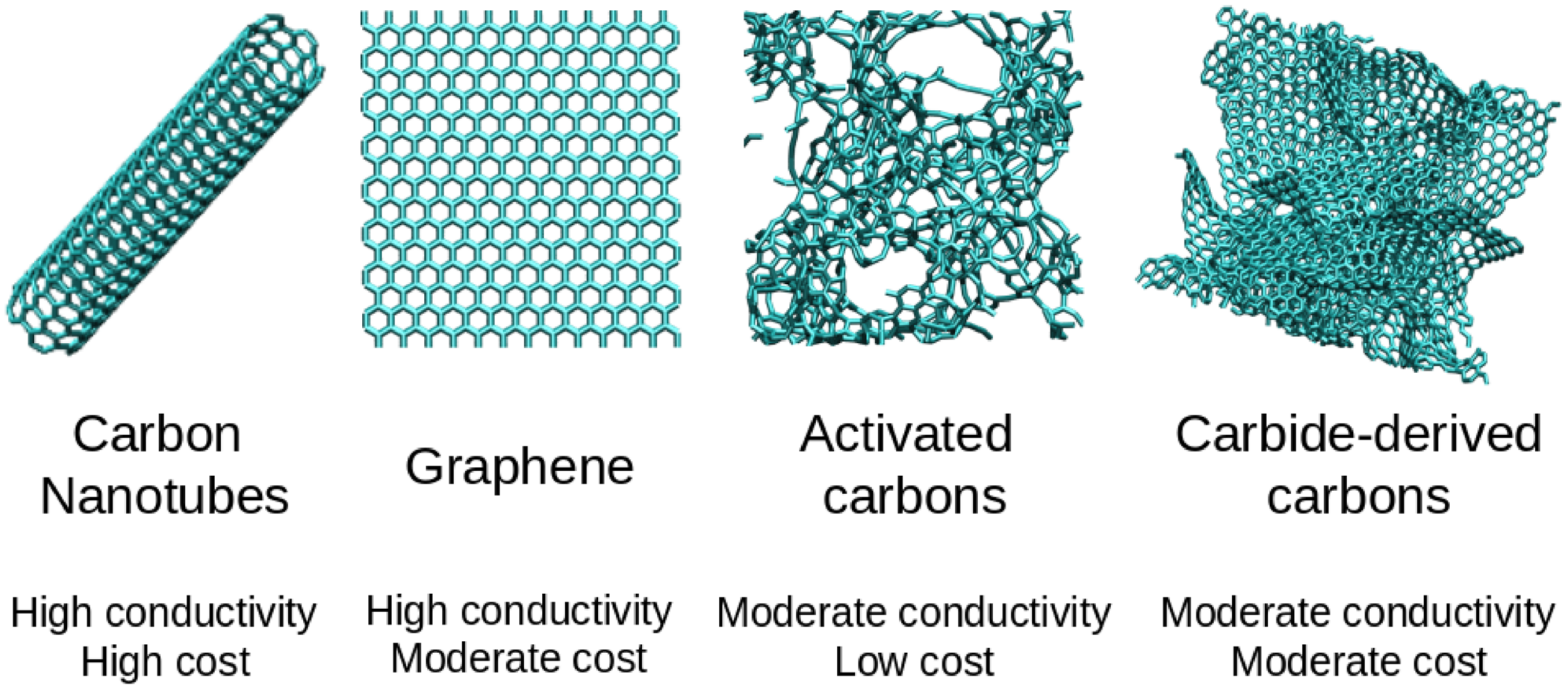

**Figure 1.** Illustration of a range of existing carbon morphologies. [Adapted with permission from P. Simon and Y. Gogotsi, Acc. Chem. Res. 46 (2013), 1094-1103. Copyright 2013 American Chemical Society].

Graphene is a one-atom-thick planar sheet of $sp^2$ bonded carbon atoms organised in a honeycomb crystal lattice. Graphene attracts a lot of interest for many applications thanks to its peculiar properties such as high heat and electronic conductivities, high mechanical and chemical stabilities, and high surface area which can reach 2670 $m^2 g^{-1}$.[7] Depending on the purity required, graphene can be synthesized in a more or less costly way.[8]

Activated carbons are prepared through an activation process using various precursors (e.g. wood, coconut shells, fruit pits, synthetic polymers) and diverse activation techniques (e.g. oxidation in water vapour, KOH, or $CO_2$).[9] The activation process usually leads to a range of pore sizes, a large accessible surface area which can exceed 2000 $m^2$ $g^{-1}$, and the presence of functional groups. Thanks to their low cost and large accessible surface areas, activated carbons are commonly used but their poorly controlled porosity and surface chemistry can be a hindrance for specific devices. One class of activated carbons that have been used in several NMR studies are polyether ether ketone (PEEK)-derived carbons (PDCs). PDCs are synthesised via a two-step process whereby the polymer is first carbonised at approximately 900 °C under inert atmosphere, before being subjected to a second activation process (using e.g., water vapour or $CO_2$) to leave a predominantly microporous structure.[10] PDCs are favourable for NMR studies as they are easily synthesised from a widely available precursor, and also because the average pore size is controllable within the 1 - 2 nm regime by altering the duration of the activation process.[11]

Carbide-derived carbons (CDCs) are produced by chlorination of carbides.[12] and may be considered as closely related to activated carbons. The nature of the precursor carbide (e.g. TiC, SiC) and the synthesis temperature as well as post-treatment methods (e.g. vacuum annealing) can be used to control the pore size distribution, the accessible surface area (usually between 1000 $m^2 g^{-1}$ and 3000 $m^2$ $g^{-1}$) and the surface chemistry of these materials. Following the importance of the precursor and process temperature, the CDCs are often named after these synthesis conditions (e.g. TiC-CDC-1000 for a CDC synthesized at 1000ºC from TiC). While they are largely amorphous, similar to activated carbons, their better-controlled porosity and the low amount of functional groups makes them useful model materials to get a fundamental understanding of molecular adsorption in porous carbons.

Templated carbons can be generated following hard-templating or soft-templating approaches.[13] In the hard-template method, carbon precursors are impregnated in a porous matrix (e.g. zeolite) and subsequently polymerized or carbonized before the removal of the template. In the soft-template method, cooperative assembly of structure-directing agents (e.g. surfactants, ionic liquids) and organic precursors is achieved in solution. The resulting carbons are usually more flexible than with hard-templating and their synthesis can be controlled using different temperatures, solvents and ionic strengths. Templated carbons are usually characterised by relatively ordered structures with large accessible surface areas and mesopores, although it is now also possible to obtain microporous carbons with this approach.[13]

Other porous carbons include kerogens[14] which are solids present in sedimentary rocks for which the carbon/hydrogen ratio, evolving with time, is a crucial property affecting, among others, structural and mechanical properties.

*2.2 Local structure*

Most of the materials described above, and in particular commercial activated carbons, are amorphous which makes their atomistic structure hard to characterize and leads to challenges in describing structure-property or structure-performance relationships in such materials. Following observations with a number of experimental techniques, it is now quite widely accepted that porous carbons contain a majority of $sp^2$-hybridised carbons and 6-membered carbon rings, but $sp^3$ carbons and odd-membered carbon rings are also present in varying amounts.[15] Figure 2 introduces some of the experimental techniques used to investigate porous carbon structures along with some of the challenges associated with such studies.

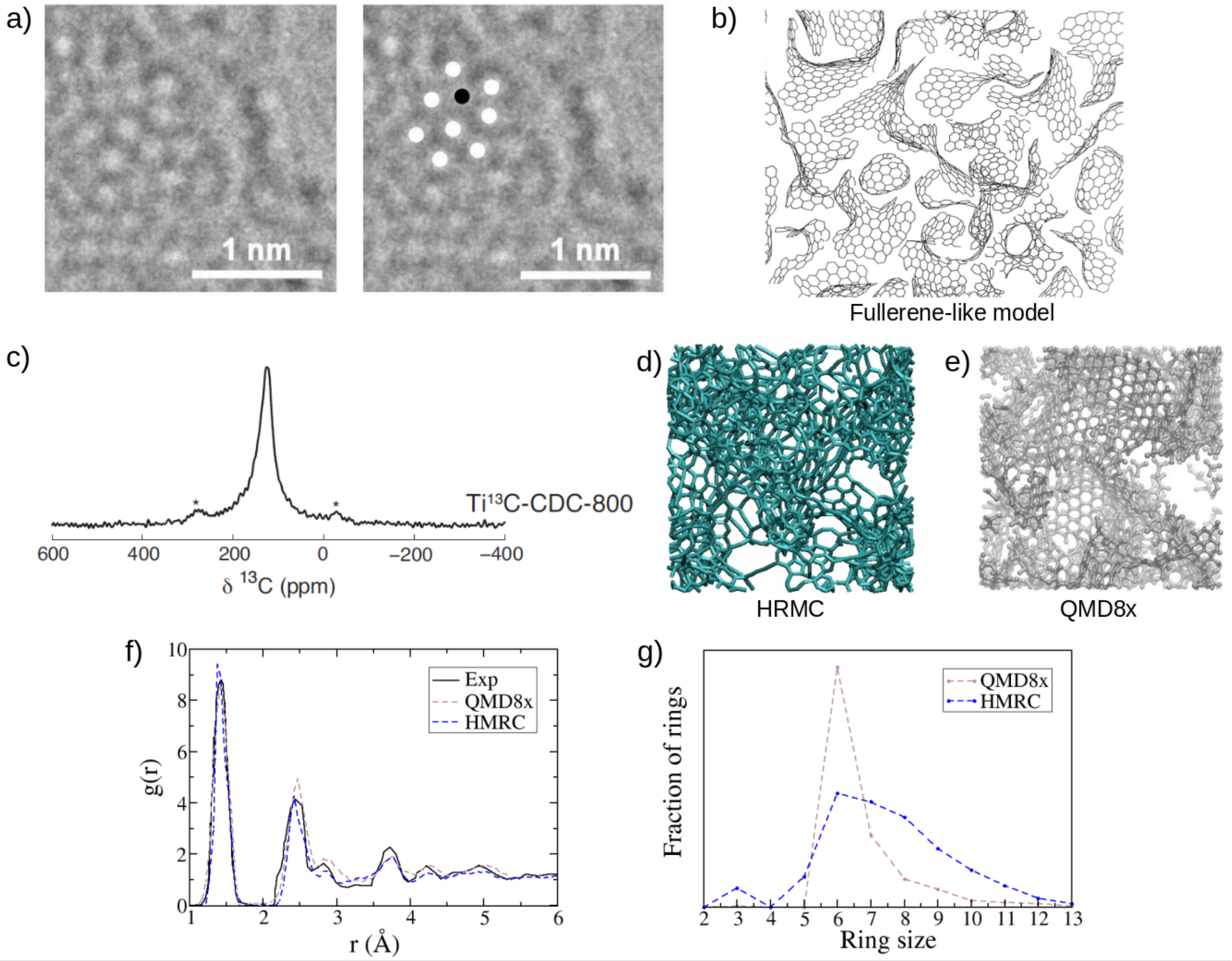

**Figure 2.** a) Aberration-corrected HRTEM micrograph of fresh activated carbon without and with white dots indicating 6-membered rings and black dot indicating possible 5-membered ring [Reproduced with permission from Ref. [16]]. b) Fullerene-like model proposed for non-graphitizing carbons [Reproduced from Ref. [15] with permission from Taylor & Francis Ltd, http://www.tandfonline.com.]. c) $^{13}C$ magic-angle spinning (MAS) NMR spectrum of TiC-CDC synthesized at 800°C. [Reproduced from Ref. [17] Copyright 2014 American Chemical Society]. d) Atomistic model of TiC-CDC-800 obtained through Hybrid Reverse Monte Carlo simulations [Reproduced from Ref. [18] with permission from The Royal Society of Chemistry]. e) Atomistic model of TiC-CDC-800 obtained through Quenched Molecular Dynamics simulations [Reproduced from Ref [19] with permission from Elsevier]. f) Pair distribution functions of the two atomistic models compared with experimental results. g) Ring statistics of the two atomistic models.

Aberration corrected high-resolution transmission electron microscopy (HRTEM) is a very valuable method to investigate the atomistic structure of microporous carbons as it can provide a direct image of the carbon rings. Harris *et al.* have used such a technique to prove the existence of 5-membered rings in activated carbons (see Figure 2a).[16] Before this work, X-ray and neutron diffraction studies had suggested the existence of 5-membered rings but without direct evidence.[20] Following the HRTEM observations, Harris *et al.* have proposed a fullerene like model to describe the atomistic structure of porous carbons (see Figure 2b).[21] This model provides adsorptive properties consistent with experimental observations.

With the chemical shifts of carbon atoms being sensitive to the local structure and coordination, $^{13}C$ NMR has been used to obtain insights into the local structure of porous carbons. Figure 2c gives the NMR spectrum obtained for a TiC-CDC synthesized at 800°C.[17] The spectrum shows a broad peak at approximately 125 ppm a value quite close to the one of graphene at around 120 ppm[22] consistent with the presence of $sp^2$-hybridised carbon atoms. For TiC-CDCs with different synthesis temperatures, the NMR spectra are quite similar limiting their use in local structure characterization.[17] In kerogens, the ratio of carbon and hydrogen atoms has a large impact on the carbon coordination and chemical shift leading to more significant variations in the NMR spectra.[23]

The determination of pair distribution functions (PDF) from X-ray or neutron diffraction experiments constitute a major source of information to characterize the local structure of amorphous carbons but these curves must be interpreted with care as a range of very different atomistic structures can all lead to average properties in agreement with experimental results. Figure 3 shows the PDF and ring statistics for a TiC-CDC synthesized at 800°C along with two atomistic models generated either by Quench Molecular Dynamics,[19] i.e. realising a quench from the liquid to the solid state of carbon in a molecular dynamics simulation, or by Hybrid Reverse Monte Carlo (HRMC),[18] i.e. obtaining an atomistic structure by random moves of the carbon positions to fit experimental data. As seen in the figure, both structures are visually different and contain very different rings but both structures provide PDF in good agreement with experimental data. This results from the one-dimensionality of the PDF which does not provide enough information to differentiate between multiple possible structures.

As a consequence of the amorphous nature and limited set of techniques able to directly image porous carbons, the local structure of these materials is qualitatively understood but quantitative and precise atomistic structures are still missing.

### *2.3 Mesoscale structure and porosity*

In addition to structural heterogeneity at the molecular level, porous carbons present a large variety at the mesoscale. The determination of the pore size distribution and accessible surface area of amorphous carbons is essential in order to understand structure-property relationships. Nevertheless, as for the atomistic structure, this is a challenge. The most common approach to determine the pore size distribution is through the measurement of adsorption isotherms using small molecules such as argon, nitrogen or carbon dioxide. Figure 3a shows the adsorption isotherm measured using $N_2$ and activated carbon fibre ACF-15.[24]

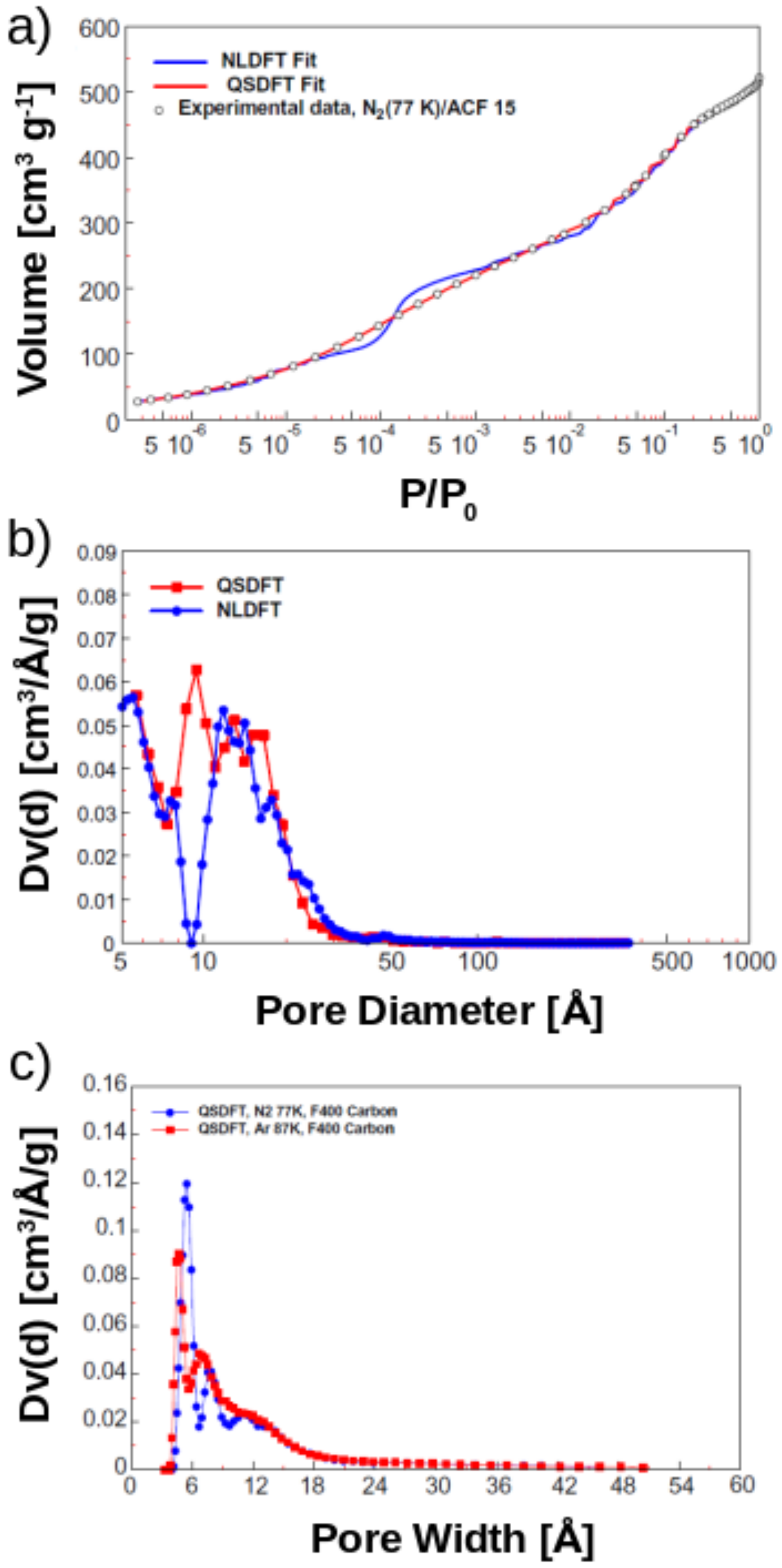

**Figure 3.** a) Experimental $N_2$ adsorption isotherm (77 K) together with the NLDFT and QSDFT theoretical isotherms for activated carbon fibre ACF-15. b) NLDFT and QSDFT differential pore size distributions for activated carbon fiber ACF-15. c) Pore size distributions of activated carbon Calgon F400 calculated with QSDFT from nitrogen and argon isotherms. All figures are reproduced from Ref. 24 with permission from Elsevier.

To extract the pore size distribution from gas adsorption isotherm data, the approach usually adopted is to decompose the experimental isotherm into a kernel of partial isotherms. This step usually employs a number of assumptions. In the Barrett-Joyner-Halenda (BJH) approach, for example, it is assumed that the pores are not connected and have a cylindrical shape, and that the surface is totally wetted by the liquid.[25] In the Non Local Density Functional Theory (NLDFT) method, the pores are also considered disconnected but very often with a slit shape with a smooth surface.[26] The Quench Solid Density Functional Theory method (QSDFT) is similar to NLDFT but includes a certain roughness of the pore surface to flatten the steps predicted by NLDFT but not present in the experimental adsorption isotherms.[24] The lack of controlled model microporous carbon materials with well-defined pore sizes makes it difficult to validate the accuracy of these methods.

Similar to the interpretation of pair distribution functions, the description of the pore size distribution from adsorption isotherms faces some challenges. The different models (e.g. QSDFT, NLDFT) or pore shapes adopted lead to different pore size distributions, and the use of different molecules for the experimental measurements also induce sometimes dramatic variations in the determined pore size distributions.[24] These effects can be seen in Figure 3b and 3c The measurement of the accessible surface area is also affected by such challenges.

### 3. Observation of adsorbed species by NMR spectroscopy

*3.1 Introduction to ring-current effects*

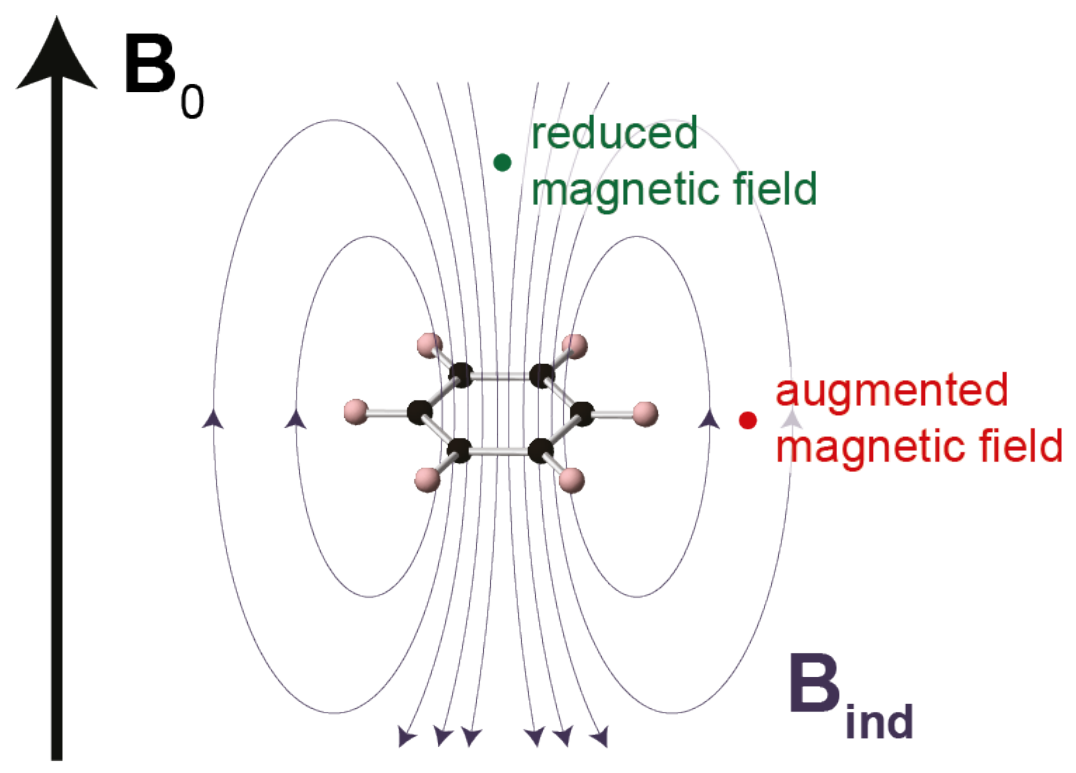

**Figure 4.** Illustration of the ring current-induced magnetic field, $\mathbf{B_{ind}}$, which results from the application of an external field, $\mathbf{B_0}$, to a π-electron system obeying Huckel's 4*n*+2 rule. Species located above (or below) the ring will be in a shielded environment, whereas species located to the side of the ring will be in a deshielded environment.

From the point of view of NMR spectroscopy, one of the most important consequences of the primarily $sp^2$-hybridised structure of porous carbons is the so-called ring current effect. This term refers to the circulation of π-electrons within conjugated carbon rings in response to an applied magnetic field. The ring current model was first developed in separate studies by Pauling[27] and Lonsdale,[28] who showed that the induced current within a six-membered ring gives rise to a secondary local magnetic field. Based on the simple classical approach taken in these early studies, the ring current-induced field for closed shell rings obeying Huckel's 4*n*+2 rule is always diamagnetic and therefore reduces the local magnetic field above and below the ring, but augments it at locations around the ring (Figure 4). As was subsequently demonstrated by Pople,[29] and Waugh & Fessenden,[30] the magnitudes of ring current-induced secondary fields are sufficient to have a measurable effect on the chemical shifts of species located within them. A classic example is paracyclophanes where the chemical shifts of bridging alkyl protons located above the plane of a six-membered ring are reduced by several ppm.[30, 31] Since the early classical treatments of ring current effects, more advanced approaches for investigating them have been developed based on density functional theory (DFT) and symmetry considerations.[32] However, in many cases simple considerations of the geometry and orientation of delocalised electron systems provide a good approximation for identifying regions of locally reduced or augmented magnetic field. On this basis, ring current induced shifts have become widely used to provide structural information about the relative positions of aromatic centres and other nearby species in solution- and solid-state NMR.[33-36]

For porous carbons, the interior carbon surfaces can be considered as extended arrangements of conjugated rings. Modelling ring currents within extended conjugated structures can be complex depending on the topology and perimeter structure, as well as any defects present. However, the magnitude of the ring current shift above a $sp^2$-hybridised carbon surface can be estimated using a simple phenomenological model. Anderson *et al.*[37] derived a relationship for the chemical shift change due to the ring current-induced field, Δδ, given by

$$\Delta\delta = \frac{(\chi^{\parallel} - \chi^{\perp})}{6\pi z^3} \quad (1)$$

where $\chi^{\parallel}$ and $\chi^{\perp}$ are the respective molar magnetic susceptibilities of the surface parallel and perpendicular to the applied field, and $z$ is the distance above the carbon surface. Taking the known value of $(\chi^{\parallel} - \chi^{\perp}) = 3.1 \times 10^{-9}$ $m^3$ $mol^{-1}$ for graphite at room temperature, a value of $\Delta\delta = -10.2$ ppm is obtained at a distance of 3 Å from the carbon surface. Due to the $z^{-3}$ dependence of the ring-current induced magnetic field, $\Delta\delta$ reduces to less than –1 ppm at a distance of ~7 Å from the surface. This simple approximation shows that the ring current-induced shift is a very localised effect, and measurable chemical shift changes should only be observed for species in very close proximity to the carbon surface, *i.e.*, at distances relevant to adsorption. Ring current shifts can therefore provide a highly selective probe of adsorbates, as they affect only species that are in very close proximity to the carbon surface. For microporous carbons, only species that are located within the micropore (and/or mesopore) network - and therefore have average locations on the order of nanometres from the carbon surface - are expected to show a measurable ring current shift.

*3.2 Observation of in-pore species in porous carbons with NMR spectroscopy*

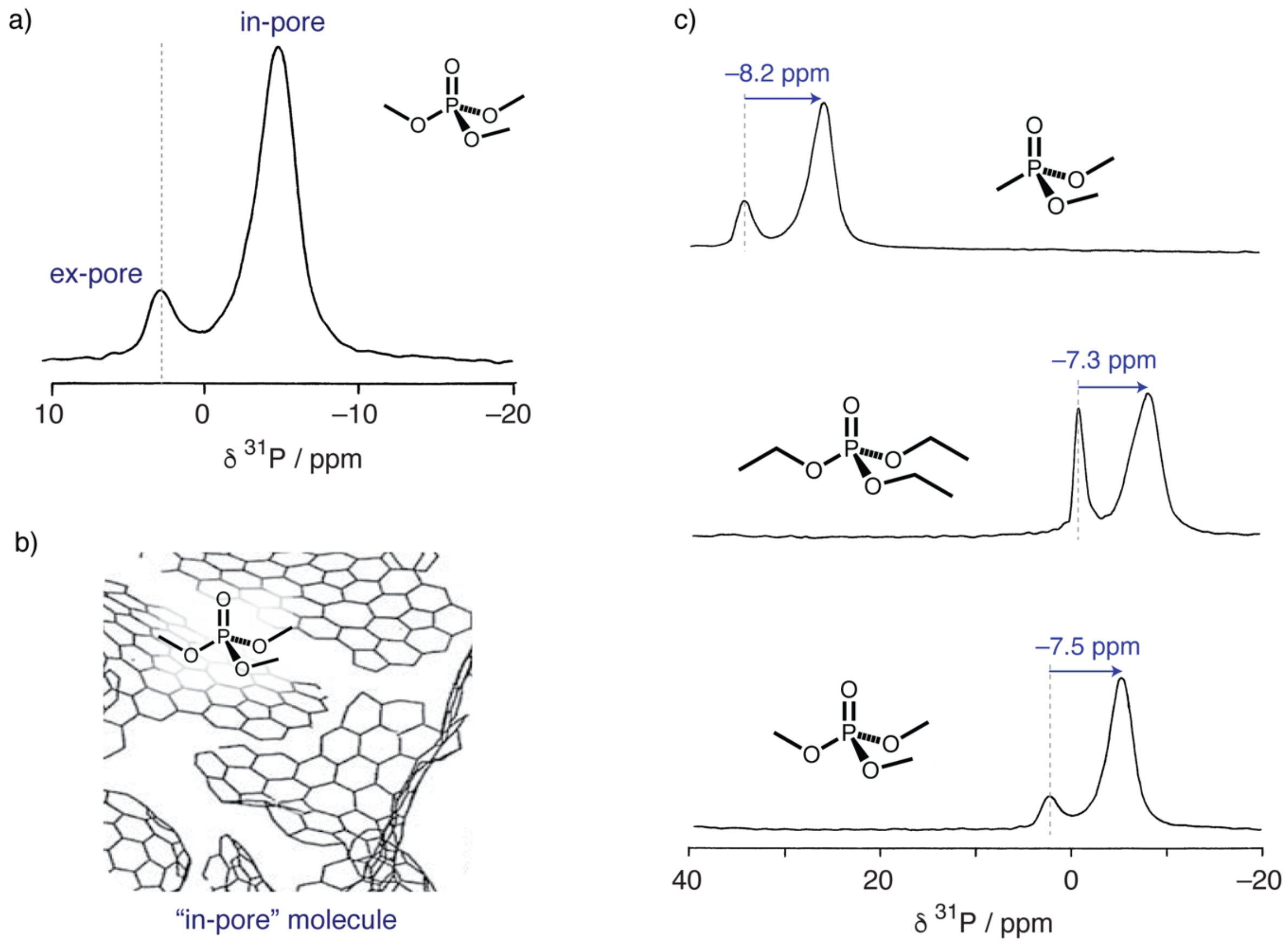

**Figure 5.** a) $^{31}P$ NMR spectrum of activated carbon soaked with trimethylphosphate. The dashed line shows the position of the $^{31}P$ chemical shift of neat liquid, and the Lewis structure of trimethlyphosphate is shown. b) Schematic showing the proposed position of in-pore molecules, nearby $sp^2$ hybridised carbon rings. [Carbon structure image reproduced from Ref. [15] with permission from Taylor & Francis Ltd, http://www.tandfonline.com] c) $^{31}P$ NMR spectra of activated carbon soaked with different phosphorous-containing organic liquids. a) and c) are adapted with permission from Ref. [38].

A series of studies from Harris and coworkers in the 1990s demonstrated how the ring current effect can be used to investigate the adsorption of liquids in activated carbons.[38-40] In these studies carbon samples were soaked with liquids and then magic-angle spinning (MAS) NMR experiments were used to study the adsorbed liquid species. Two main resonances were observed in such experiments (**Figure 5a**): (i) a resonance with a chemical shift very similar to that of the neat liquid (hereafter referred to as "ex-pore"), and (ii) a resonance shifted to lower chemical shifts assigned to species inside the carbon micropores ("in-pore").[38] It was proposed that the shift of the in-pore resonance compared to the neat liquid is primarily due to local magnetic fields associated with aromatic rings in the carbon host (**Figure 5b**), *i.e.* ring current effects. Ex-pore species are more remote from carbon surfaces and likely occupy large pores or spaces between the carbon particles. When $^{31}P$ NMR experiments were performed on a series of different organic liquids, each adsorbed on the same carbon material in separate experiments, similar chemical shift changes were observed between the in-pore and ex-pore resonances in each case (**Figure 5c**).

*3.3 Effect of varying the adsorbate for the same adsorbent*

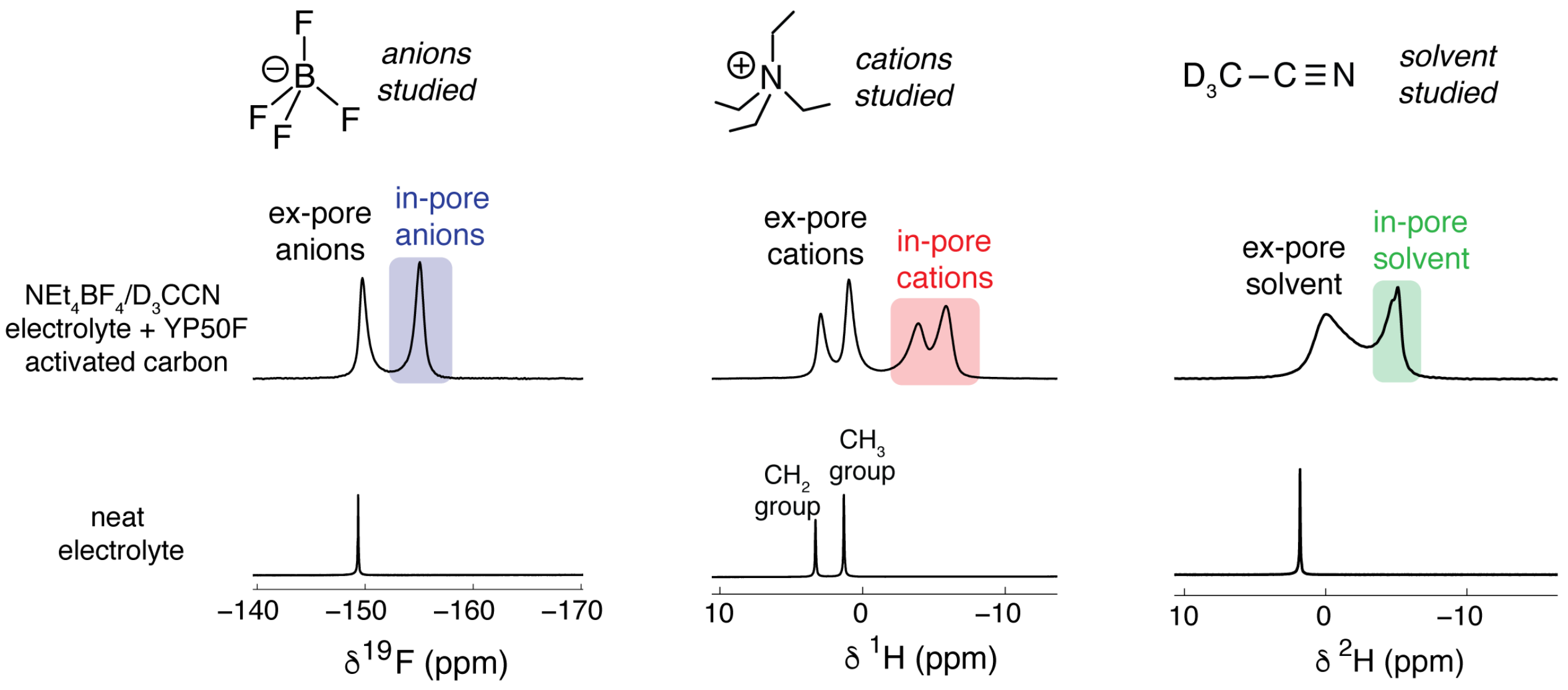

**Figure 6.** MAS NMR spectra of YP50F activated carbon film soaked with $NEt_4BF_4$ (1.5 M) in $CD_3CN$ electrolyte. The MAS rate was 5 kHz. Adapted with permission from A.C. Forse, C. Merlet, J.M. Griffin and C.P. Grey, J. Am. Chem. Soc. 138 (2016), 5731-5744. Copyright 2016 American Chemical Society.

Later NMR experiments supported the idea that the chemical shift changes due to adsorption are largely independent of the molecule and nucleus studied for a given carbon. For example NMR experiments were carried out to study the adsorption of the electrolyte $NEt_4BF_4$ (1.5 M in deuterated acetonitrile, $CD_3CN$) in activated carbon.[41] The different components of the electrolyte were investigated separately by studying different NMR nuclei ($NEt_4^+$ : $^1H$ NMR, $BF_4^-$ : $^{19}F$ NMR, $CD_3CN$ : $^2H$ NMR), and in each case in-pore and ex-pore resonances were observed (**Figure 6**). When comparing these spectra it is useful to define $\Delta\delta$ (ppm) = $\delta_{in\text{-}pore} - \delta_{neat}$, which gives a measure of the change of the chemical shift due to adsorption. $\Delta\delta$ has a value of –5.7, –7.2 and –6.9 ppm for $^{19}F$, $^1H$, and $^2H$ NMR, respectively (note there are two distinct $^1H$ resonances for the cations, and both have the same $\Delta\delta$ value here) (**Table 1**). The $\Delta\delta$ values fall within 1.2 ppm of each other, supporting the idea that the shift due to adsorption, $\Delta\delta$, largely originates from ring current effects, which are independent of the molecule and nucleus studied to a first approximation. This idea is further corroborated by collating literature NMR data for a range of systems (see **Table 1)** which shows that, for a given carbon adsorbent material, only relatively minor differences are observed between the $\Delta\delta$ values for different adsorbed molecules.

Similar $\Delta\delta$ values for different adsorbed species in a given carbon support the idea that these shifts can be modelled largely as a nucleus independent chemical shift (NICS) (see later). Some small differences are however observed between the $\Delta\delta$ values for different species in a given carbon. This suggests that (i) there are differences in the average position of the studied nuclei relative to the carbon pore walls and/or (ii) there are other chemical shielding effects beyond ring currents that arise from non-covalent interactions between the adsorbent and the adsorbate, changes in solvation or other effects.[42] An extreme example is given by an NMR study of adsorption of aqueous alkali metal halide electrolytes in PDCs, which revealed larger variations in $\Delta\delta$ values (**Table 1**). More work must be done to understand the variations in $\Delta\delta$ values for different adsorbents in a given carbon.

**Table 1.** Δδ values for MAS NMR studies where different liquids were adsorbed in common porous carbon materials. Additional sample details: [a]6 mg carbon film and ~4 µL electrolyte, [b]electrolyte volumes were matched to pore volumes obtained from gas sorption measurements, [c]3 mg carbon and 5 µL ionic liquid. [d]loadings were 60% weight/weight liquid/carbon, [e]PEEK-derived carbon with 46% burn off was used, and samples were saturated with liquid.

| Carbon | Adsorbed Liquid | Studied nucleus | Δδ (ppm) | Ref. |
|---|---|---|---|---|
| YP50F | $NEt_4BF_4$ (1.5 M) in $CD_3CN$[a] | $^{19}F$ ($BF_4^-$) | –5.7 | [41] |
| | | $^{1}H$ ($NEt_4^+$, $CH_2$) | –7.2 | |
| | | $^{1}H$ ($NEt_4^+$, $CH_3$) | –7.2 | |
| | | $^{2}H$ ($CD_3CN$) | –6.9 | |
| | $NEt_4BF_4$ (1.0 M) in $CD_3CN$[b] | $^{11}B$ ($BF_4^-$) | –6.9 | [43] |
| | | $^{1}H$ ($NEt_4^+$, $CH_2$) | –7.5 | |
| | | $^{1}H$ ($NEt_4^+$, $CH_3$) | –7.5 | |
| | | $^{2}H$ ($CD_3CN$) | –6.7 | |
| | $Pyr_{13}$TFSI ionic liquid[c] | $^{19}F$ ($TFSI^-$) | –6.6 | [44] |
| | EMI TFSI ionic liquid[c] | $^{19}F$ ($TFSI^-$) | –6.3 | |
| | EMI $BF_4$ ionic liquid[b] | $^{11}B$ ($BF_4^-$) | –6.9 | [43] |
| TiC-CDC-1000 | $NEt_4BF_4$ (1.5 M) in $CD_3CN$[a] | $^{19}F$ ($BF_4^-$) | –5.4 | [41] |
| | | $^{1}H$ ($NEt_4^+$, $CH_2$) | –5.8 | |
| | | $^{1}H$ ($NEt_4^+$, $CH_3$) | –5.7 | |
| | | $^{2}H$ ($CD_3CN$) | –5.5 | |
| | $NEt_4BF_4$ (1.0 M) in $CH_3CN$[b] | $^{11}B$ ($BF_4^-$) | –6.2 | [45] |
| TiC-CDC-800 | $NEt_4BF_4$ (1.5 M) in $CD_3CN$[a] | $^{19}F$ ($BF_4^-$) | –4.0 | [41] |
| | | $^{1}H$ ($NEt_4^+$, $CH_2$) | –4.7 | |
| | | $^{1}H$ ($NEt_4^+$, $CH_3$) | –4.6 | |
| | | $^{2}H$ ($CD_3CN$) | –3.6 | |
| TiC-CDC-600 | $NEt_4BF_4$ (1.5 M) in $CD_3CN$[a] | $^{19}F$ ($BF_4^-$) | –2.6 | [41] |
| | | $^{1}H$ ($NEt_4^+$, $CH_2$) | –4.0 | |
| | | $^{1}H$ ($NEt_4^+$, $CH_3$) | –3.9 | |
| | | $^{2}H$ ($CD_3CN$) | –3.4 | |
| | $NEt_4BF_4$ (1.0 M) in $CH_3CN$[b] | $^{11}B$ ($BF_4^-$) | –3.7 | [45] |
| ASZM[d] | trimethylphosphate | $^{31}P$ | –6.1 | [38] |
| | triethylphosphate | | –5.3 | |
| | dimethylmethylphosphonate | | –5.6 | |
| SC2[d] | trimethylphosphate | $^{31}P$ | –7.9 | [38] |
| | triethylphosphate | | –7.3 | |
| | dimethylmethylphosphonate | | –7.4 | |
| PDC[e] | $H_2O$ | $^{1}H$ | –6.4 | [46] |
| | LiCl (1 M) in $H_2O$ | $^{7}Li$ ($Li^+$) | –5.9 | |
| | NaCl (1 M) in $H_2O$ | $^{23}Na$ ($Na^+$) | –6.3 | |
| | RbCl (1 M) in $H_2O$ | $^{87}Rb$ ($Rb^+$) | –10.1 | |
| | CsCl (1 M) in $H_2O$ | $^{133}Cs$ ($Cs^+$) | –10.7 | |

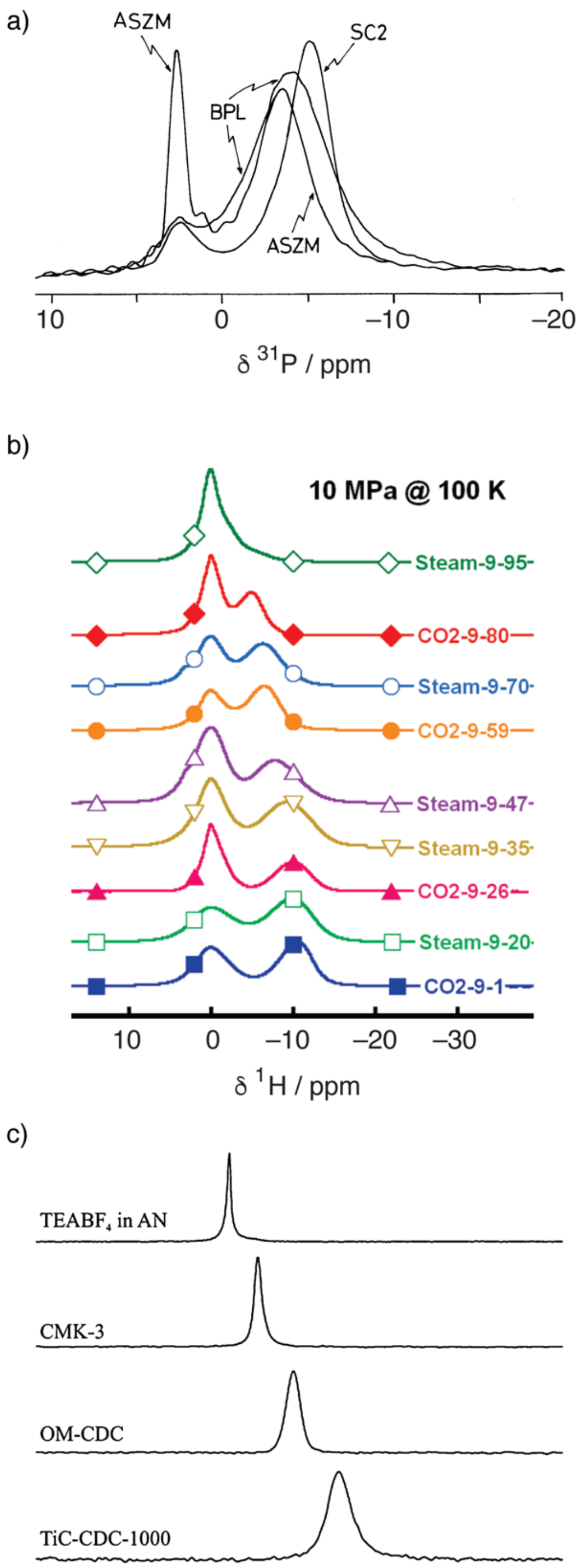

**Figure 7.** a) $^{31}$P MAS NMR spectra of 60 wt% triethylphosphate adsorbed on three different porous carbons. [Reproduced with permission from Ref. 38] b) $^{1}$H static NMR spectra of $H_2$ gas at 10 MPa and 100 K in the presence of PDCs synthesised under various conditions. [Reprinted with permission from R.J. Anderson, T.P. McNicholas, A. Kleinhammes, A. Wang, J. Liu and Y. Wu, *J. Am. Chem. Soc.* 132 (2010), 8618-8626. Copyright 2010 American Chemical Society]. c) $^{11}$B MAS NMR spectra of (top) 1 M $NEt_4$-$BF_4$ / ACN, and (below) the same electrolyte adsorbed on various porous carbons. [Reproduced with permission from Ref. 45]

### *3.4 Effect of varying the adsorbent with the same adsorbate*

**Table 1** shows that changing the carbon adsorbent for a given adsorbed molecule has a large effect on the measured $\Delta\delta$ values. This makes sense, given that ring current effects are the major factor contributing to the $\Delta\delta$ values – different carbon adsorbents will have different local structures and in turn different ring currents and chemical shielding effects for adsorbed species. Early experiments by Harris *et al.* on the adsorption of organic liquids showed that different carbons gave rise to different $\Delta\delta$ values (**Figure 7a**).[38] Similar effects were observed in studies of $H_2$ and $H_2O$ adsorption in a series of PDCs (**Figure 7b**),[37, 47] and later in studies of electrolyte adsorption in supercapacitor electrode materials (**Figure 7c**).[17, 45, 46, 48] These findings suggest that it should be possible to obtain structural information about a given carbon adsorbent through analysis of the NMR spectra of adsorbed species, a topic that is discussed further in Section 5.

*3.5 Experimental considerations for the observation of in-pore species*

One of the key requirements in studying adsorption in porous carbons is to be able to sufficiently resolve the in-pore and ex-pore resonances to allow accurate measurement of $\Delta\delta$ and/or quantify the resonance intensities. In general, the local mobility of in-pore and ex-pore species is sufficient to average both dipolar and chemical shift anisotropy interactions. For many systems, linewidths are sufficiently narrow that the in-pore and ex-pore species can be resolved under both MAS and static conditions, although static linewidths tend to be larger. The choice between MAS and static measurements is therefore largely dictated by the nature of the investigation being performed: MAS experiments are generally preferable where possible due to the increased resolution; however, static measurements have been favoured in many *in situ* investigations of the effect of charging the carbon surface (as discussed in Section 7), or experiments that vary the adsorbate gas pressure.

The reasons for the resolution difference between static and MAS measurements have not been systematically investigated. However, a likely contribution is local and bulk magnetic susceptibility effects, which are known to be a major source of line broadening in multiphase samples.[49] Indeed, it is notable that ex-pore resonances for saturated samples are generally significantly broader than resonances for the same neat liquid, showing that the presence of the carbon particles in the ex-pore liquid has a marked effect on the linewidth. In addition to the linewidths, susceptibility effects in static oriented samples can also have measurable effects on the observed shift. Static measurements on porous carbons are often carried out on carbon films formed into flat plate or disc shapes. Forse *et al.* demonstrated that changing the orientation of a porous carbon freestanding film electrode caused shift changes of ~10 and ~8 ppm for the in-pore and ex-pore resonances, respectively (Figure 8).[50] This highlights the need for careful consideration of sample geometry and orientation when rationalising the magnitude of $\Delta\delta$ in static measurements, particularly when comparing values between different studies.

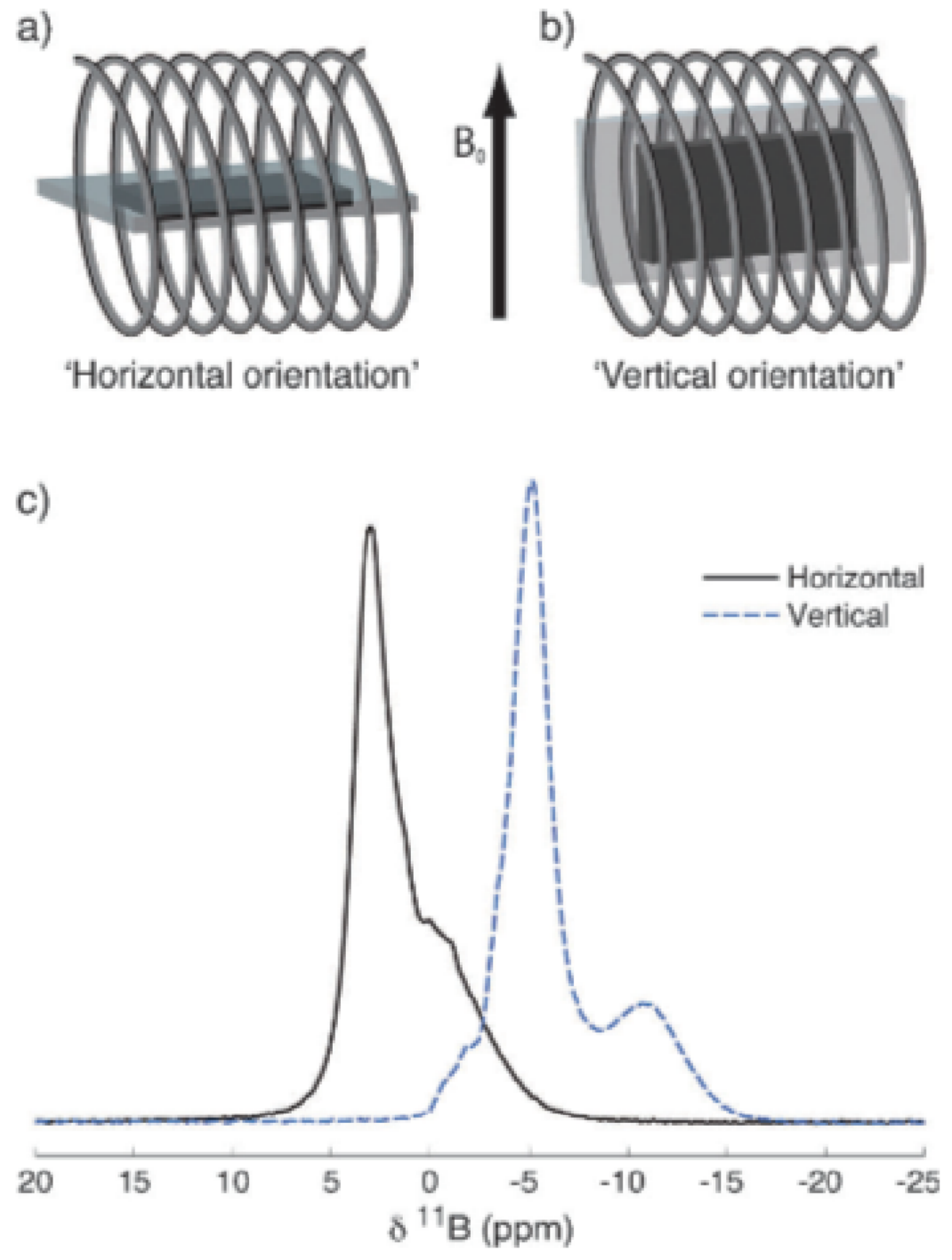

**Figure 8.** Illustrations of (a) horizontal and (b) vertical carbon film sample orientations. (c) The resulting static $^{11}B$ NMR spectra of a piece of TiC- CDC-1000 film soaked with 10.0 mL of $NEt_4BF_4$/ACN electrolyte. [Reproduced from Ref. 50 with permission from the PCCP Owner Societies]

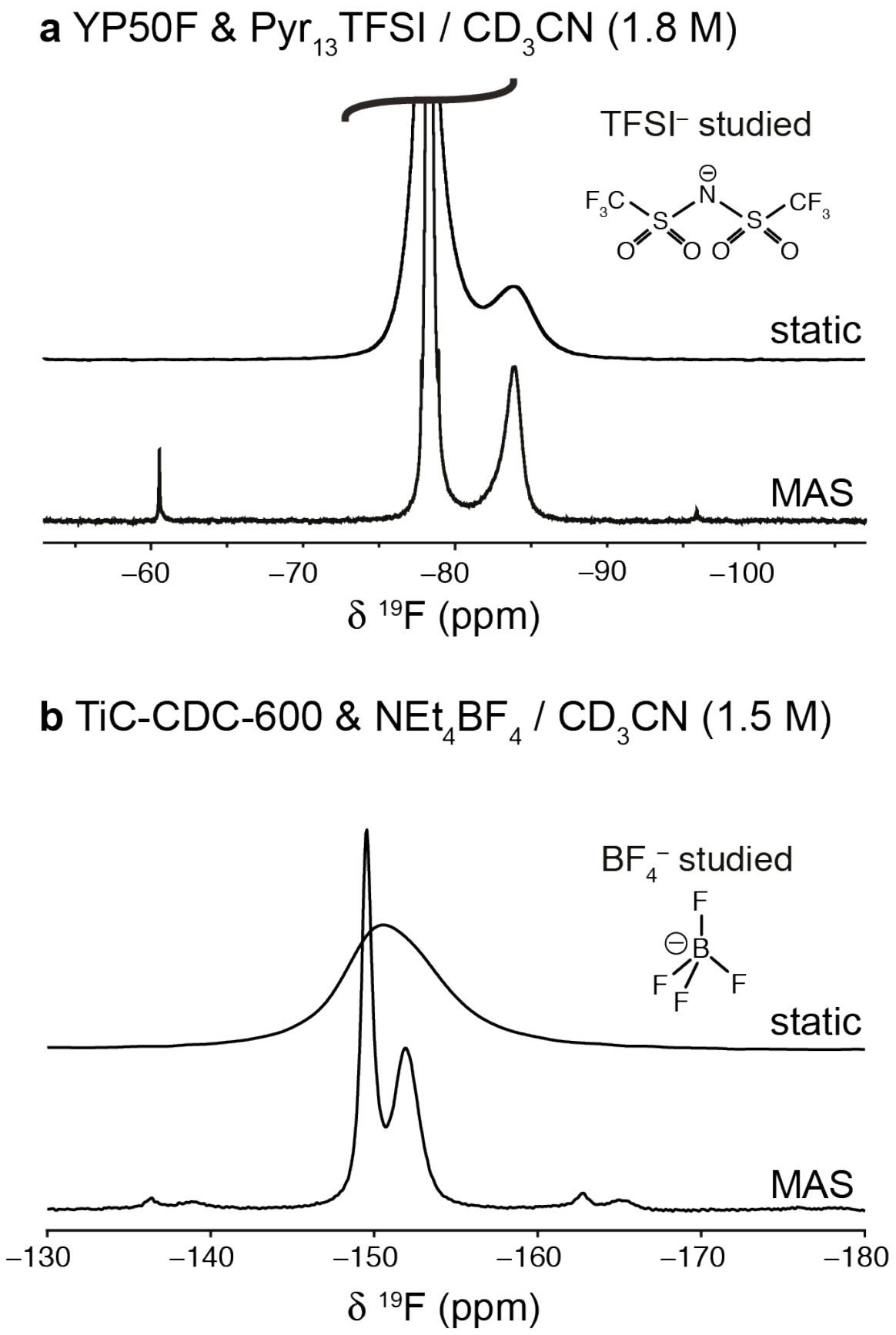

**Figure 9.** Comparison of $^{19}F$ static and MAS NMR spectra for two different carbon electrolyte systems: (a) ($Pyr_{13}TFSI$) / $CD_3CN$ adsorbed on YP50F and (b) $NEt_4BF_4$ / $CD_3CN$ adsorbed on TiC-CDC-600. For (b), the in-pore and ex-pore resonances are only resolved when MAS is used. Adapted from Ref. 41.

Another contribution to the linewidth is the range of local environments occupied by the in-pore species. The amorphous structures of porous carbons mean that the interior pore space is made up of a range of different pore widths, geometries and local aromaticities, even for carbons with highly uniform pore size such as TiC-CDCs. Using two-dimensional exchange experiments with short mixing times, Alam & Osborn-Popp,[51] and more recently Cervini *et al.*,[46] have shown that much of the in-pore linewidth can be attributed to inhomogeneous broadening consistent with the adsorbed species occupying a range of pore environments. The extent of the in-pore line broadening will depend upon both the range of environments present within the pore structure as well as the rate at which in-pore species exchange between them (a subject that is discussed in more detail in Section 6). This can have important consequences for the choice of experimental protocol. As shown in Figure 9, for 1-Methyl-1-propylpyrrolidinium bis(trifluoromethanesulfonyl)imide ($Pyr_{13}TFSI$) / $CD_3CN$ adsorbed on YP50F, the ex-pore and in-pore environments are resolved in both static and MAS spectra meaning that either approach can be used. In contrast, for $NEt_4BF_4$ / $CD_3CN$ adsorbed on TiC-CDC-600, the two environments are unresolved in the static spectrum, meaning that MAS must be used. In this case, the structural disorder of TiC-CDC-600 and the reduced diffusion kinetics within the narrow pore network likely both contribute to the lower resolution in the static measurement, while broadening associated with paramagnetic impurities can also not be ruled out. YP50F, on the other hand, has a larger average domain size suggesting increased structural order,[48] and also has a small proportion of mesopores that facilitate fast diffusion through the pore network, meaning that the in-pore and ex-pore are often resolved in static measurements. Nevertheless, for highly viscous adsorbates with slow diffusion kinetics such as neat ionic liquids, significant broadening of the in-pore resonance is still observed, meaning that static measurements may not be possible.[44]

## 4. Theoretical investigations of adsorption-induced chemical shift changes

*4.1 Nucleus-independent chemical shift.*

As discussed in the previous section, consideration of ring current effects can help to explain the diamagnetic shift often observed for adsorbed species. However, it can be difficult to separate the ring current-induced contribution to $\Delta\delta$ from chemical, dynamic, or other magnetic effects, even when systematic measurements are carried out using the same carbon or adsorbate. One method to determine the contribution of the ring current-induced shielding to $\Delta\delta$ is through the theoretical calculation of the nucleus-independent chemical shift (NICS).[52] In the NICS approach, DFT calculations are used to determine the electronic structure of a model carbon surface. This model can be a fragment, extended $\pi$-conjugated molecule, or a periodic structure depending on the DFT method used. Once the electronic structure has been calculated (usually following geometry-optimisation of the model structure), the full magnetic response can be determined. This approach is essentially the same as for the calculation of NMR parameters that is commonly used for spectral interpretation and analysis; however, in the NICS approach, rather than calculating NMR parameters at nuclear positions within the model structure, the chemical shielding tensors and their isotropic components ($\sigma_{iso}$) are determined at defined points in space within or near the carbon surface. The isotropic NICS at that position is then given by $-(\sigma_{iso} - \sigma_{ref})$, where the reference shielding, $\sigma_{ref}$, is equal to zero, i.e., NICS = $-\sigma_{iso}$. The NICS value represents the change in local magnetic field that any nucleus located at the defined position would experience due to the ring currents in the nearby carbon surface. Importantly, because the calculation does not actually place an atom at the defined position, the NICS does not include any contributions from chemical or electrostatic interactions between the probe species and the carbon surface. In this way, the shielding effect due to the magnetic response of the carbon surface can be isolated from other effects that may be present in a real system.

*4.2 Applications of the NICS for the study of aromaticity.*

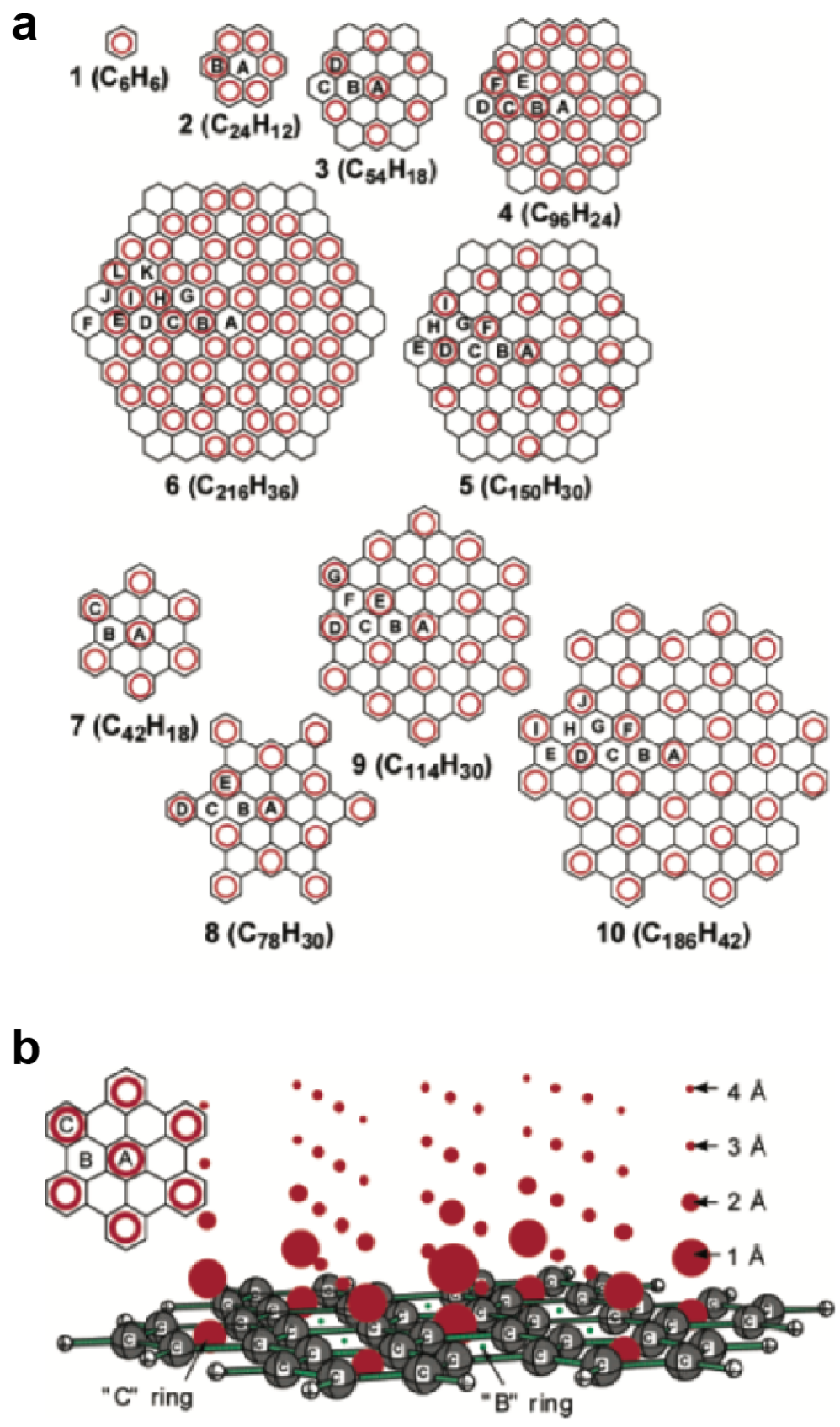

**Figure 10.** (a) Aromatic rings within PBHs (indicated by red circles) identified on the basis of strongly diatropic NICS values calculated at ring centres. (b) Calculated NICS values at various positions above a PBH surface with aromatic and non-aromatic rings (as represented in the inset). The magnitude of the NICS (represented by the size of the red circles) tends towards an average value at a distance of 4 Å [Adapted with permission from D. Moran, F. Stahl, H.F. Bettinger, H.F. Schaefer and P.v.R. Schleyer, J. Am. Chem. Soc. 125 (2003), 6746-6752. Copyright 2003 American Chemical Society].

Although the concept of the NICS was proposed as far back as 1958, its calculation by quantum chemical means was pioneered by Schleyer & coworkers in the 1990s for the study of aromaticity in π-conjugated structures. Through the study of a number of five-membered ring heterocycles, it was found that the magnitude of the NICS at the ring centre shows a strong linear correlation with the aromatic stabilisation energy of the surrounding ring.[53] Therefore, the NICS provides a convenient parameter by which to indirectly explore aromaticity as it captures the often complex magnetic response of the delocalised electrons in conjugated systems as a single parameter defined at a point in space. Moran *et al.* later applied the NICS approach to a series of extended polybenzoid hydrocarbons based on six-membered rings.[54] Despite the apparent structural simplicity and symmetry of some of the systems studied, the aromaticity in extended π-conjugated structures was found to be complex and dependent upon the nature of the perimeter structure (Figure 10a). Large NICS values indicative of aromaticity were only found for alternating rings which obey the Clar sextet rule,[55] whereas adjacent six-membered rings showed small diatropic or weakly paratropic NICS indicating they were not aromatic. However, by calculating the NICS at points in space above the ring planes, it was found that by a distance of approximately 4 Å, NICS values above aromatic and non-aromatic rings converge to an average value (Figure 10b). This highlights that the NICS must be determined close to the ring centre in order to probe local aromaticity, but also that at distances relevant to molecular adsorption (≥ 4 Å) much of the complexity of the aromaticity within a graphene-like surface is lost and adsorbate species in the vicinity of a carbon surface are predicted to experience an average ring current-induced shielding.

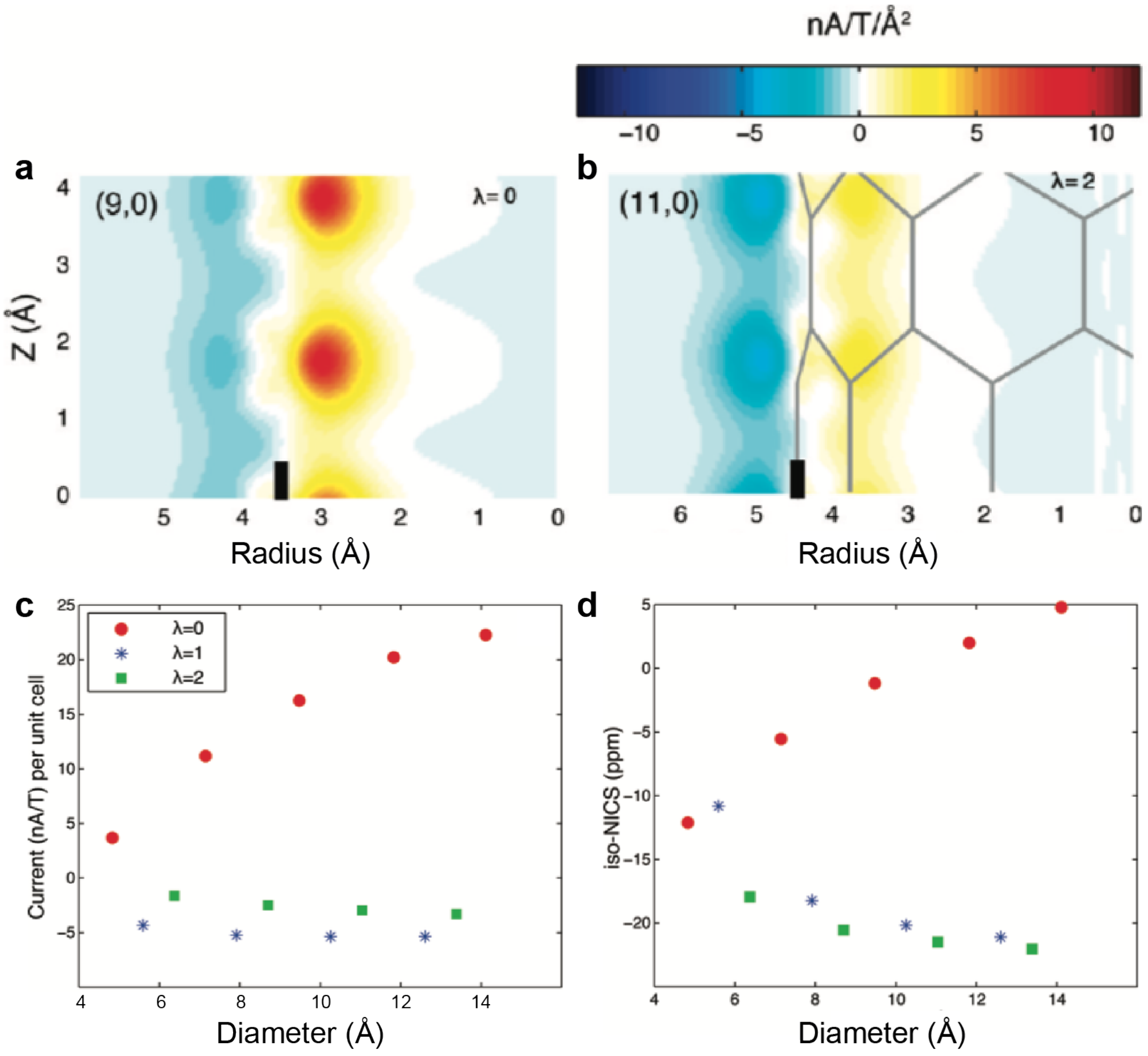

**Figure 11.** (a) A map of the induced current flow in a (9,0) nanotube showing positive (paratropic) current flow on the inner surface and negative (diatropic) current flow on the exterior surface (the nanotube edge is indicated by the black line) (b) The induced paratropic current is reduced while the induced diatropic current is increased for a (11,0) nanotube of different chirality. (c) Calculated total induced current and (d) calculated isotropic NICS as a function of nanotube diameter for three families of nanotube chirality (indicated by the parameter λ). [Adapted with permission from M. Kibalchenko, M.C. Payne and J.R. Yates, ACS Nano 5 (2011), 537-545. Copyright 2011 American Chemical Society].

*4.3 Theoretical studies of the NICS in carbon nanotubes.*

The theoretical work by Moran *et al.* provides valuable insight into the NICS within and near model carbon fragments; however, the models studied consisted of a single carbon surface. For adsorbates within microporous carbons, the majority of the adsorbed species are confined within partially enclosed carbon nanostructures. Insight into the shift changes for species in such environments can be obtained by considering theoretical studies on nanotubes, which provide a well-defined geometry that can be systematically varied in terms of pore size (diameter) and aromaticity (chirality). Kibalchenko *et al.* used a periodic DFT approach to explicitly map the magnetically-induced current as well as the NICS within infinite zigzag nanotubes.[42] Interestingly, the induced current maps were found to be complex, with many nanotubes showing both diatropic (or aromatic) and paratropic (or anti-aromatic) behaviour in different regions of the tube surface (Figures 11a,b). Different relative contributions of the local currents to the total induced current resulted in either a net paramagnetic or diamagnetic response which depended on the chirality of the nanotube. The total induced current was found to correlate well with the *z*-component of the NICS tensor but the isotropic NICS varied in a more complex way, in some cases resulting in a net diamagnetic shift for nanotubes with an overall positive induced current. The relationship between the isotropic NICS and the nanotube diameter was also non-trivial, where the magnitude of the NICS either increased or reduced with increasing diameter, depending on the chirality of the nanotube (Figures 11c,d). This shows that while nanotubes may have conceptually simple structures, their magnetic response can be complex and requires detailed consideration of the bonding topology.

**Table 2.** Calculated change in isotropic chemical shift, Δδ, for species within a (9,0) carbon nanotube. The corresponding isotropic NICS for the same nanotube is shown in the right-hand column. [Adapted with permission from M. Kibalchenko, M.C. Payne and J.R. Yates, ACS Nano 5 (2011), 537-545. Copyright 2011 American Chemical Society].

| | | $CH_4$ | | HCl | | |
|---|---|---|---|---|---|---|
| | He | H | C | H | Cl | NICS (ppm) |
| Δδ (ppm) | –5.0 | –4.3 | –3.6 | –5.0 | 4.9 | –5.5 |

**Table 3.** Calculated change in isotropic chemical shift in ppm for an encapsulated HCl molecule compared to the isotropic NICS for nanotubes of varying diameter. [Adapted with permission from M. Kibalchenko, M.C. Payne and J.R. Yates, ACS Nano 5 (2011), 537-545. Copyright 2011 American Chemical Society].

| | | HCl | | |
|---|---|---|---|---|
| | Diameter (Å) | H | Cl | NICS (ppm) |
| (10,0) | 7.9 | –18.0 | –13.7 | –18.2 |
| (11,0) | 8.7 | –20.4 | –20.0 | –20.6 |
| (12,0) | 9.5 | –1.8 | –2.1 | –1.2 |
| (13,0) | 10.3 | –20.1 | –20.5 | –20.2 |

The work by Kibalchenko *et al.* also revealed that the chemical shift change (i.e., the Δδ value) of encapsulated species can differ significantly from that predicted by the NICS alone. A series of probe species were placed within a 7.1 Å diameter nanotube and the calculated chemical shifts were compared to the isotropic NICS of –5.5 ppm calculated for the same empty nanotube (Table 2). The calculated chemical shift of an encapsulated He atom (–5 ppm) was very similar to the calculated NICS. However, a Cl atom within an encapsulated HCl molecule showed a positive (apparently paramagnetic) chemical shift change of 4.9 ppm. In contrast, the chemical shift of the H atom in the same molecule (–5 ppm) was again very similar to the isotropic NICS. Further calculations showed that the deviation between the calculated shift and the NICS was largest for small diameter nanotubes, suggesting that the effect arises from an interaction between the Cl atom and the tube walls (Table 3). This highlights that there can be additional contributions to Δδ values that are not accounted for by the NICS alone. Possible explanations in this particular case include (1) a physical interaction between the Cl atom and the nanotube walls resulting in a perturbation of the Cl electron distribution, (2) a chemical interaction between the Cl atom and the nanotube involving a partial or complete charge transfer, or (3) a local disruption of the ring currents in the nanotube surface caused by the presence of the Cl atom resulting in a local change in the NICS. Without further study, it is difficult to separate these possible contributions to $\Delta\delta$, and it is also possible that each may contribute to some degree. However, recent DFT calculations have shown that strong interactions between Cl atoms and nanotubes can take place.[56] In particular, the adsorption of a Cl atom in a nanotube leads to a spontaneous charge transfer and the formation of a $Cl^-$ species. In this study, the chemical shifts were not assessed but such a charge transfer would be expected to have a large impact on the NMR parameters.

In a DFT study of methanol within carbon nanotubes by Ren *et al.*, the calculated NICS for $^{17}O$ within the methanol molecule differed from the calculated chemical shift. This difference was attributed to weak non-bonding interactions with the nanotube surface, which polarises the π-electron cloud within the nanotube wall, therefore locally modifying the magnetic environment of the probe nucleus.[57] For highly polarisable atoms, chemical shift changes can arise directly from polarisation of the electron cloud of the atom itself; *e.g.*, for xenon atoms adsorbed in porous carbons, the dominant contributions to the $^{129}Xe$ shift are atom-atom and atom-surface interactions, rather than the NICS.[58] Recent experimental work on aqueous electrolyte adsorption has also shown that $\Delta\delta$ for heavy ions such as $Rb^+$ and $Cs^+$ adsorbed in microporous carbons can deviate significantly that predicted from the NICS - although in this case desolvation effects are also a possible factor (Table 1).[46] Further experimental and computational work is required to properly understand and separate the contributions to chemical shift changes for adsorbed species so that $\Delta\delta$ values can be more accurately interpreted to obtain structural and dynamic information.

*4.4 Modelling of carbon slit pores.*

While nanotubes provide a well-defined and conceptually simple model for exploring enclosed carbon structures, real microporous carbons rarely show such long-range structural homogeneity. Although it is very difficult to accurately model the disorder that is present in microporous carbons, slit-shaped pores are generally considered to be a closer representation of the structure than the cylindrical pore model implied by nanotubes. A simple structural model for a slit pore geometry could be two infinite graphene sheets separated by a defined distance. However, the vanishing band gap at the Dirac point in the electronic structure of graphene can complicate the calculation of magnetic properties in periodic DFT codes, leading to spurious results.[59] In view of this, molecular-based DFT approaches have been used, where the graphene sheets are approximated by polycyclic molecules of a defined size. Forse *et al.* have explored the variation of the NICS with the size and separation of model carbon molecules based on coronenes.[17] For the smallest model surface studied, coronene, a NICS value of –4.5 ppm was calculated at 3 Å from a single surface, but this reduced to almost 0 ppm at a distance of 15 Å (Figure 12a). This is consistent with the strong distance dependence predicted by the phenomenological model in Eq. 1, and confirms that measurable Δδ should only be expected for species with average locations in close proximity to the carbon surface. The effect of the domain size of the model carbon surface was found to strongly influence the calculated NICS, with the value at 3 Å increasing to –8 ppm for the larger dicircumcoronene molecule (Figure 12b). This highlights that the local structure within a microporous carbon may also strongly affect the magnitude of the ring current-induced shift experienced by adsorbed species; *i.e.*, carbons with larger average domain sizes may be expected to give larger Δδ values for similar proximities of adsorbate to the carbon surface. In other studies, molecular size effects have been studied for a wide range of coronenes showing an interesting pattern with the number of ring shells at short distances while the trend at larger distances (above 4Å) is monotonous with the molecular size.[48, 60-62]

Within a slit pore formed by two parallel carbon fragments, the effects of the NICS were found to be approximately additive, with the NICS at any location between the sheets being essentially equal to the sum of the NICS from both surfaces (Figure 12c). Furthermore, to properly model the expected ring current-induced shift, the effects of fast molecular diffusion within the pore should be accounted for by averaging the calculated NICS profile. For narrow pores, this gave larger average NICS values of up to *e.g.*, –10.6 ppm for an 8 Å pore between two circumcoronene sheets. As the separation between the carbon surfaces was increased, the average NICS reduced due to the increasing region of space with negligible NICS. For the largest separation studied of 40 Å, an average NICS value of –3.7 ppm was predicted. This shows that while the ring current shift reduces with increasing pore width, measurable effects can still be expected even for species adsorbed within relatively large micropores and small mesopores.

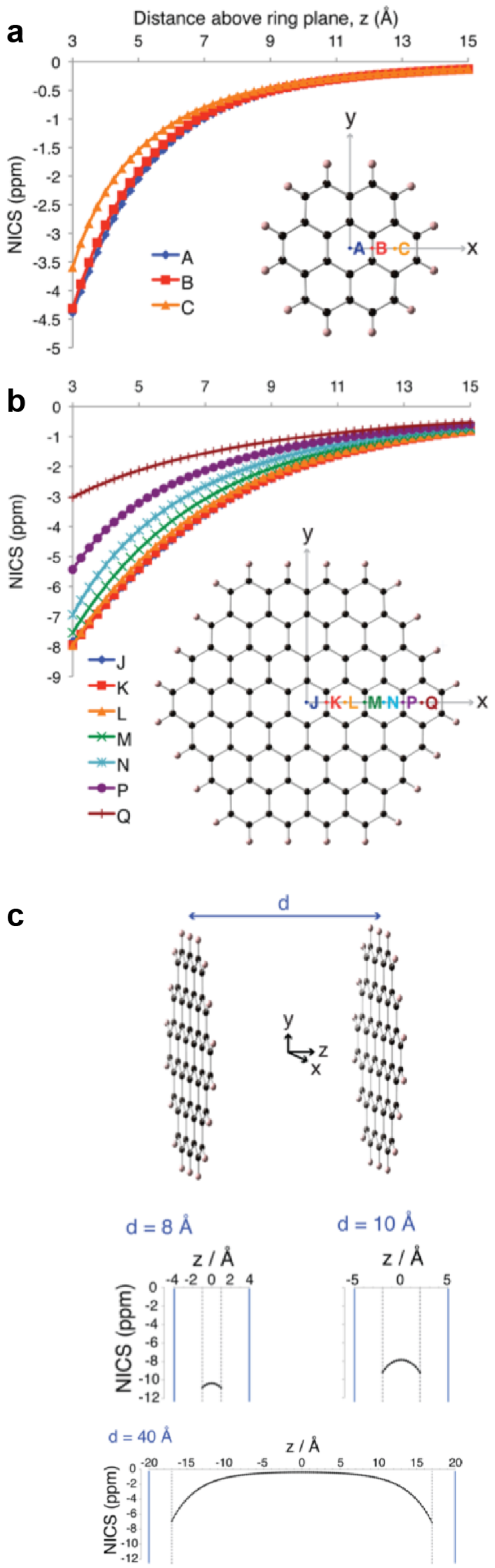

**Figure 12.** Calculated NICS profiles above (a) coronene and (b) dicircumcoronene at different positions across the surface. (c) representation of a slit pore formed by two parallel circumcoronene molecules and plots of the NICS profiles for different inter-plane separations. [Adapted with permission from A.C. Forse, J.M. Griffin, V. Presser, Y. Gogotsi and C.P. Grey, *J. Phys. Chem. C* 118 (2014), 7508-7514. Copyright 2014 American Chemical Society].

*4.5 Effects of local curvature and defects.*

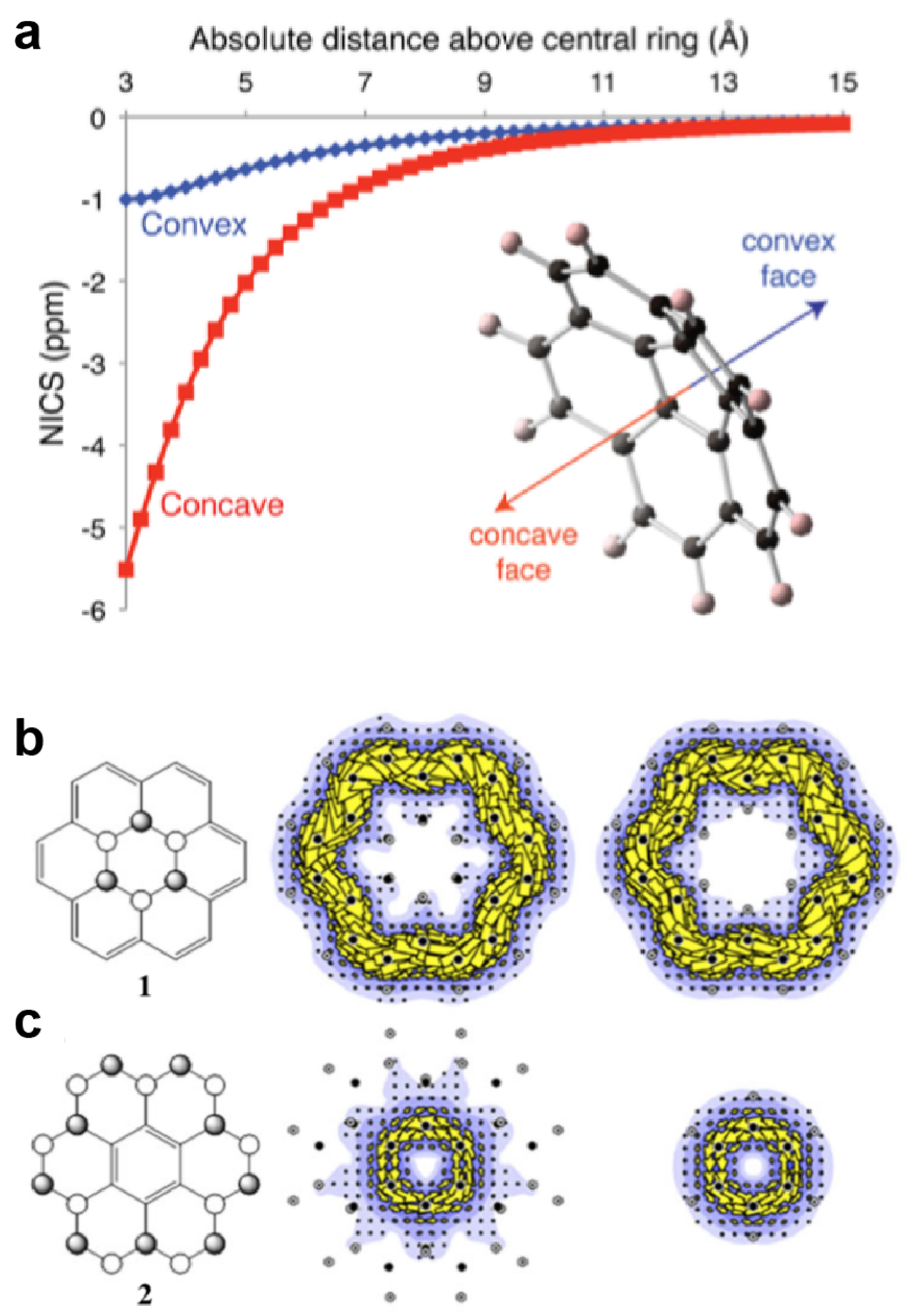

**Figure 13.** (a) Calculated NICS profile for corannulene. [Adapted with permission from A.C. Forse, J.M. Griffin, V. Presser, Y. Gogotsi and C.P. Grey, J. Phys. Chem. C 118 (2014), 7508-7514. Copyright 2014 American Chemical Society]. (b, c) Hydrogenation of coronene as indicated by light (pointing down) and dark (pointing up) circles (left colum), and the corresponding ring current density maps calculated at the ipsocentric CHF/6-31G** level (middle column). The right-hand column shows the ring current map for annulene and benzene [Reproduced with permission from P.W. Fowler, C.M. Gibson and D.E. Bean (2014) Writing with ring currents: selectively hydrogenated polycyclic aromatics as finite models of graphene and graphene *Proc. Roy. Soc.* A **470** 20130617 https://doi.org/10.1098/rspa.2013.0617].

Although slit pores represent a simple and reasonable approximation for the overall pore geometry within microporous carbons, it is important to recognise that most real microporous carbons have very little defined structure beyond the length scale of a few carbon rings, and generally contain extensive curvature and defects. As discussed in Section 2, it has been suggested that local curvature is associated with the presence of non-six membered rings which disrupt the planarity of the structure. Such defects could have non-negligible effects on the NICS near the carbon surface. Merino *et al.* have shown that anti-aromatic rings such as $C_4H_4$ give rise to significant paratropic ring currents, which would be expected to paramagnetically shift species located above the ring system.[63] Buhl also observed large paratropic NICS values at non-aromatic ring centres in fullerenes and curved carbon fragments.[64] For corannulene, the ring-centre NICS of the central five-membered ring was calculated to be 6 ppm (paratropic) whereas the values for the surrounding six-membered rings were –9 ppm (diatropic). However, as discussed in Section 4.2, it is important to note that at distances relevant to molecular adsorption (≥ 3 Å) averaging of the NICS values for individual rings can occur. Indeed, for the fullerenes considered by Buhl, endohedral NICS values evaluated at the centres of mass were found to be diatropic, despite the presence of paratropic rings within the fullerene surface. Forse *et al.* also modelled NICS profiles perpendicular to the surface of corannulene, and similarly found net diatropic NICS on both sides of the molecule (Figure 13a).[17] The maximum NICS at a distance of 3 Å from the concave face was –5.5 ppm whereas the maximum NICS for the convex face was –1 ppm. These values can be compared with maximum NICS of –4.5 ppm for planar coronene, which falls between these two values. This result shows that while regions of local carbon curvature may not lead to a paratropic NICS at distances relevant to molecular adsorption, porous carbons containing different amounts of concave, convex, and planar surfaces may result in different chemical shifts for adsorbed species (though the effects of dynamic averaging within the pore space should again be considered).

In addition to local curvature, chemical defects may also be expected to influence the local aromaticity and therefore the NICS experienced by adsorbed species. One source of defects is at domain edges, which can be terminated with hydrogen or other functional groups. The study by Forse *et al.* also showed that the calculated NICS above the carbon surface reduces near the edges of coronenes. Maximum NICS values above perimeter rings were approximately –3 ppm for the structures studied, compared to a value of –8 ppm above the central ring of dicircumcoronene. However, it is also important to note that in real materials, adsorbed species can in principle explore all parts of the structure which are physically accessible, and this may include locations beside domain edges. As discussed in Section 3, the local magnetic field beside aromatic rings should be augmented by the ring current effect and so adsorbed species will be deshielded in these locations. This has not been studied extensively, but DFT calculations predict a small positive NICS near the ring edge of coronene in plane with the molecule.[41] For structures with a high concentration of domain edges, such a deshielding effect may partially counteract the diamagnetic shielding in locations perpendicular to domain surfaces when dynamic averaging is taken into account. Indeed, several experimental studies on TiC-CDCs have reported smaller $\Delta\delta$ for species adsorbed in small-pore samples than in large-pore samples.[17, 45] This cannot be rationalised on the basis of the pore-size dependence of the NICS (which predicts the opposite effect) but can instead be interpreted by the higher degree of disorder and smaller domain size in smaller pore TiC-CDCs.[19]

Chemical defects can also be present within carbon domains in the form of hydrogen atoms or other functional groups chemisorbed to the surface. Species bonded to the carbon surface may disrupt the aromaticity and hence ring currents due to the formation of $sp^3$- rather than $sp^2$-hybridised carbon atoms. Fowler *et al.* have examined the effect of partial hydrogenation of small carbon domains by DFT methods.[65] It was found that for small numbers of hydrogenous defects, the resulting ring current mimicked that expected for the resulting electron delocalisation that was formed (Figure 13b,c). For example, hydrogenation of the central ring of coronene leaves a delocalised electron system around the perimeter of the molecule resembling that of annulene. Ring current calculations on hydrogenated coronene and annulene were then found to give very similar results. Similarly, hydrogenation of the perimeter of coronene leaves a central delocalised electron system and ring currents similar to that in benzene. This work suggests that although surface defects within porous carbons can locally disrupt the aromaticity, longer-range ring currents can be maintained providing a suitable conjugation pathway exists. The NICS was not explicitly calculated in this work but would be expected to reflect the ring currents. However, further work is required to determine the precise effect of surface defects on the NICS and the density of defects that is required to disrupt the long-range ring currents pathways.

## 5. Quantifying Carbon Structure by NMR Spectroscopy

### *5.1 Quantifying Carbon Pore Size Distributions from Ring Current Shifts*

As discussed in Sections 3 and 4, chemical shift changes due to adsorption ($\Delta\delta$) are dominated by ring current effects, which strongly depend on the structure of the carbon adsorbent. Variations in $\Delta\delta$ values for a given adsorbate molecule in a series of different carbon adsorbates have been rationalised based on differences in carbon pore sizes,[45-47] and differences in the local ordering of the graphene-like fragments of the carbon.[48] These findings suggest it should be possible to go one step further and obtain structural information for carbon materials by measuring NMR spectra for adsorbed molecules to determine $\Delta\delta$ values, which can be related to the carbon structure via NMR calculations. Going from the NMR spectrum to structural information requires a connection to be made between the measured $\Delta\delta$ values and carbon structures. This step presents the biggest challenge in determining structural information by this approach, as many different structural features affect $\Delta\delta$ values (pore size, structural order, curvature etc.) which all must be accounted for.

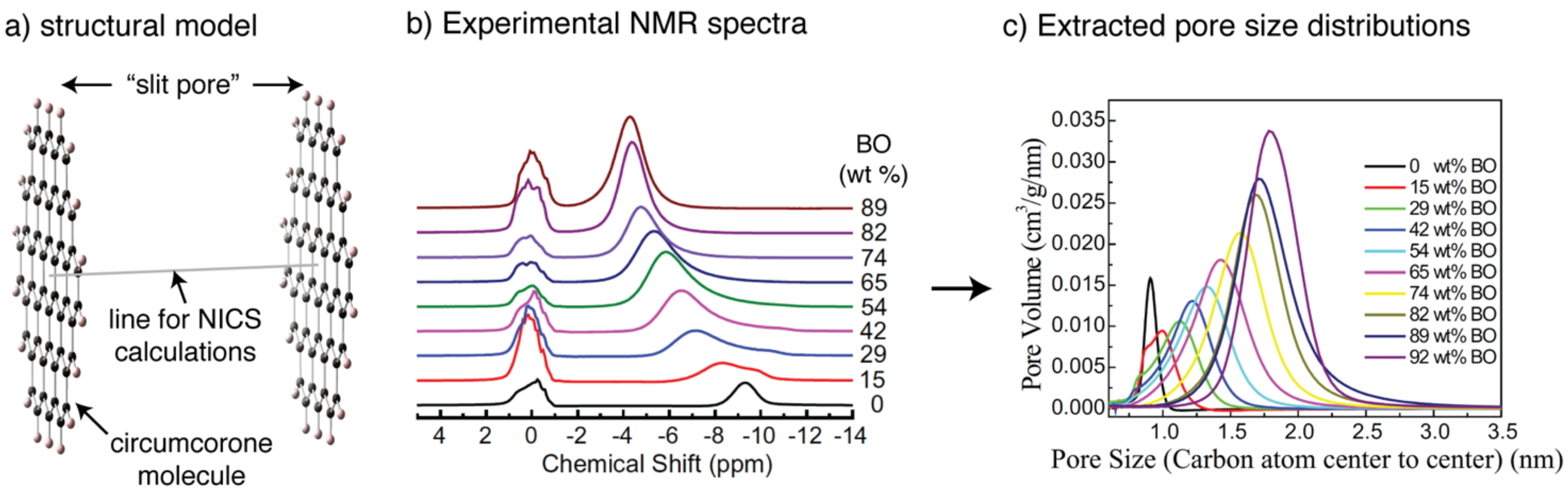

**Figure 14.** a) Slit pore model with circumcoronene "walls" used for NICS calculations. This is used to establish a connection between pore size and Δδ values. b) Experimental $^1$H MAS NMR spectra for a series of PDCs saturated with water. The different carbons have different burn-off (BO) levels – higher BO means more material was removed (burned off) during the carbon synthesis. c) Pore size distributions extracted from the NMR data in b). b) and c) are reproduced with permission from Ref. [47].

Xing *et al*. developed a promising approach to extract carbon pore size distributions from the $^1$H NMR spectra of adsorbed water.[47] Their approach involves the measurement of the $^1$H MAS NMR spectrum of water adsorbed in a carbon sample saturated with water, followed by the conversion of this spectrum to a pore size distribution. This conversion is achieved by making the following assumptions: (i) the carbon system can be modelled as a series of slit pores, where the pore walls are circumcoronene molecules (Figure 14a), (ii) the Δδ value for the adsorbed water can be modelled solely as a NICS which is calculated using DFT along a line between the centres of the two circumcoronene molecules (Figure 14a), (iii) water molecules undergo fast exchange between different positions along this line and are equally likely to be at any point along this line (with positions within 3 Å of the circumcoronene atoms forbidden). With these three assumptions there is a unique mapping of a given pore size to a NICS value. Finally it is assumed (iv) that the experimental NMR spectra are in the slow exchange regime with regards to the exchange of water molecules between pores of different sizes. Using these assumptions the spectral intensity at a given chemical shift is converted to a pore volume for a given pore size. The authors analysed a series of PDCs with different structures and pore size distributions were obtained (Figures 14b,c).

The validity of the assumptions in this model should be explored in future studies. Notably the choice of circumcoronene to represent the structure of the pore wall is somewhat arbitrary, and the selection of other polyaromatic hydrocarbon molecules in the NICS calculations strongly impacts the obtained NICS values and therefore pore size distributions. One may expect that different structural models may also be needed to model different carbon materials. Moreover, preferential water adsorption may occur at carbon surfaces, as observed in molecular simulations,[66-68] suggesting that a simple unweighted average of the calculated NICS may lead to errors. In their study, Xing *et al*. did not validate their NMR-derived pore size distributions via a secondary method, making it challenging to assess the validity of the obtained pore size distributions. In a later study, Alam & Osborn Popp applied the NMR method to a porous carbon soaked with a carbonate solvent.[51] Importantly they compared the NMR-derived pore size distribution to that obtained by a secondary method – classical density functional theory analysis of nitrogen and argon adsorption isotherms (a state-of-the-art approach). The two pore size distributions showed some clear differences, with a unimodal pore size distribution obtained with the NMR method, and a bimodal pore size distribution obtained via the gas sorption method. The hypotheses used in the approach developed by Xing *et al.* indeed prevent the suggestion of a bimodal distribution. A further study by Cervini *et al.* found that average pore sizes determined by the NMR method and gas sorption analysis showed reasonable agreement suggesting that the NMR method is promising.[46] Overall, more work must be done to establish the validity of the NMR approach and to explore the underlying assumptions in the analysis.

*5.2 Quantifying Carbon Order from Ring Current Shifts*

Using a conceptually related approach, Forse *et al.* used an NMR method to estimate representative carbon domain sizes for a series of porous carbon materials.[48] In this method experimental $\Delta\delta$ values are first measured for a series of carbons. A lattice simulation method is then used to calculate theoretical $\Delta\delta$ values for a series of structural models. The key features of these simulations are (i) a slit pore model is used to represent individual carbon pores (Figure 15a), (ii) the adsorption thermodynamics of molecules in the slit-pore are accounted for by inputting free energies from molecular dynamics simulations, (iii) NICS values in the slit pore are calculated using DFT, (iv) a distribution of pore sizes is input into the simulation to exactly match the experimental pore size distribution from gas sorption analysis, (v) an NMR simulation is performed in the fast exchange limit to obtain a simulated $\Delta\delta$ value. Separate simulations are run in which the carbon domain size is varied between different polyaromatic hydrocarbons (Figure 15a), and $\Delta\delta$ values are calculated.

Simulated $\Delta\delta$ values are then compared to experimental values to estimate representative carbon domain sizes (Figure 15b). For example, TiC-CDC-800 has an experimental $\Delta\delta$ value which is comparable to the simulated value for coronene domains – this suggests the carbon material has carbon domains with a size similar to that of a coronene molecule. According to this analysis, carbons that were vacuum annealed at 1400 ºC have much larger domains, comparable in size to dicircumcoronene. Some qualitative validation of the results was obtained by measurement of pair distribution functions of the studied carbons.

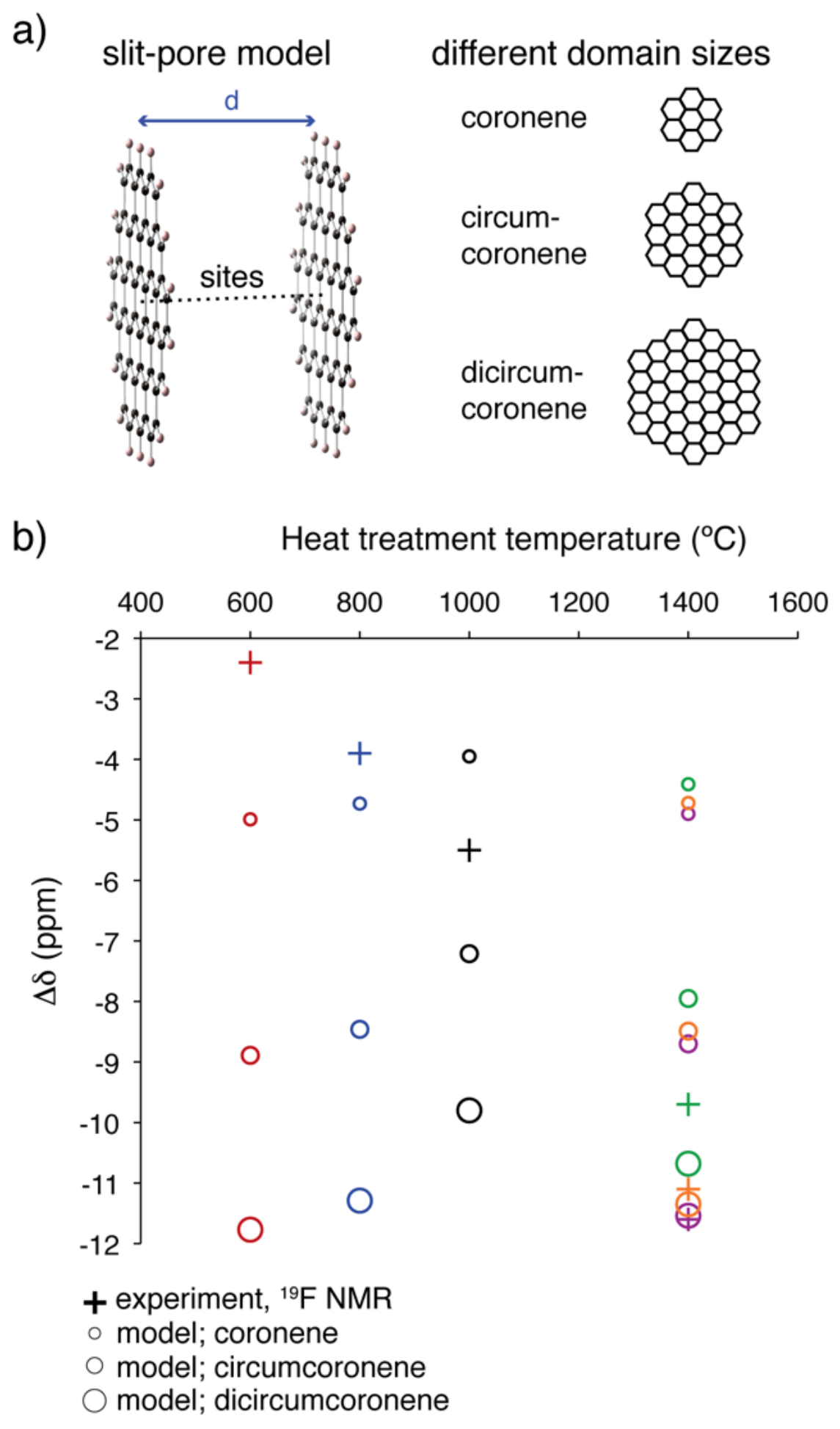

**Figure 15.** a) A slit pore model used to estimate carbon domain sizes from experimental Δδ values. Separate simulations are performed in which the domain size is varied. b) Experimental and simulated Δδ values for $^{19}F$ NMR studies of $BF_4^-$ ions adsorbed in a series of different carbons. Experimental values can be compared with the different simulated values to estimate a representative carbon domain size. [Adapted with permission from A.C. Forse, C. Merlet, P.K. Allan, E.K. Humphreys, J.M. Griffin, M. Aslan, M. Zeiger, V. Presser, Y. Gogotsi and C.P. Grey, Chem. Mater. 27 (2015), 6848-6857. Copyright 2015 American Chemical Society].

A key assumption underlying this approach, as well as the above approach to determine pore size, is that a given carbon can be well represented by a single characteristic domain size with a slit pore model. In reality porous carbons have highly heterogeneous structures and these models are highly oversimplified. Moreover, Δδ values are accounted for solely using the NICS part, but other factors may contribute to the Δδ values as discussed in Section 4. In future studies, work must be carried out to better simulate NMR spectra for molecules adsorbed in more realistic carbon models and to better establish the connection between Δδ values and carbon structure.

*5.3 Quantifying Carbon Pore Volume by NMR Measurements of Adsorbed Species*

The pore volume of a carbon material is an important quantity that will determine its adsorption properties. Carbon pore volumes are typically determined via measurement of gas adsorption isotherms using nitrogen or argon. Recently it has been shown that NMR spectroscopy experiments provide an alternative method to quantify carbon pore volumes. This is achieved via the integration of in-pore resonances for carbon samples that have been saturated with a liquid adsorbate, so that it can be assumed that all accessible pore volume is occupied by adsorbate molecules. With appropriate calibration, from the integrated in-pore intensity one can obtain the number of adsorbed moles. This number is then divided by the fluid density (mol cm$^{-3}$) to afford a pore volume. A key challenge with this approach is that the fluid density in the carbon pores is difficult to determine and assumptions are generally made when choosing a value. In spite of this challenge, Xing *et al.* found that NMR-derived carbon pore volumes (from $^{1}$H NMR of adsorbed water) for PDCs showed very good agreement with pore volumes determined by nitrogen gas sorption experiments.[47] In a later study Cervini *et al.* also found a similar good agreement between pore volumes measured by $^{1}$H NMR of PDC samples saturated with water and by nitrogen gas sorption measurements (Figure 16).[46]

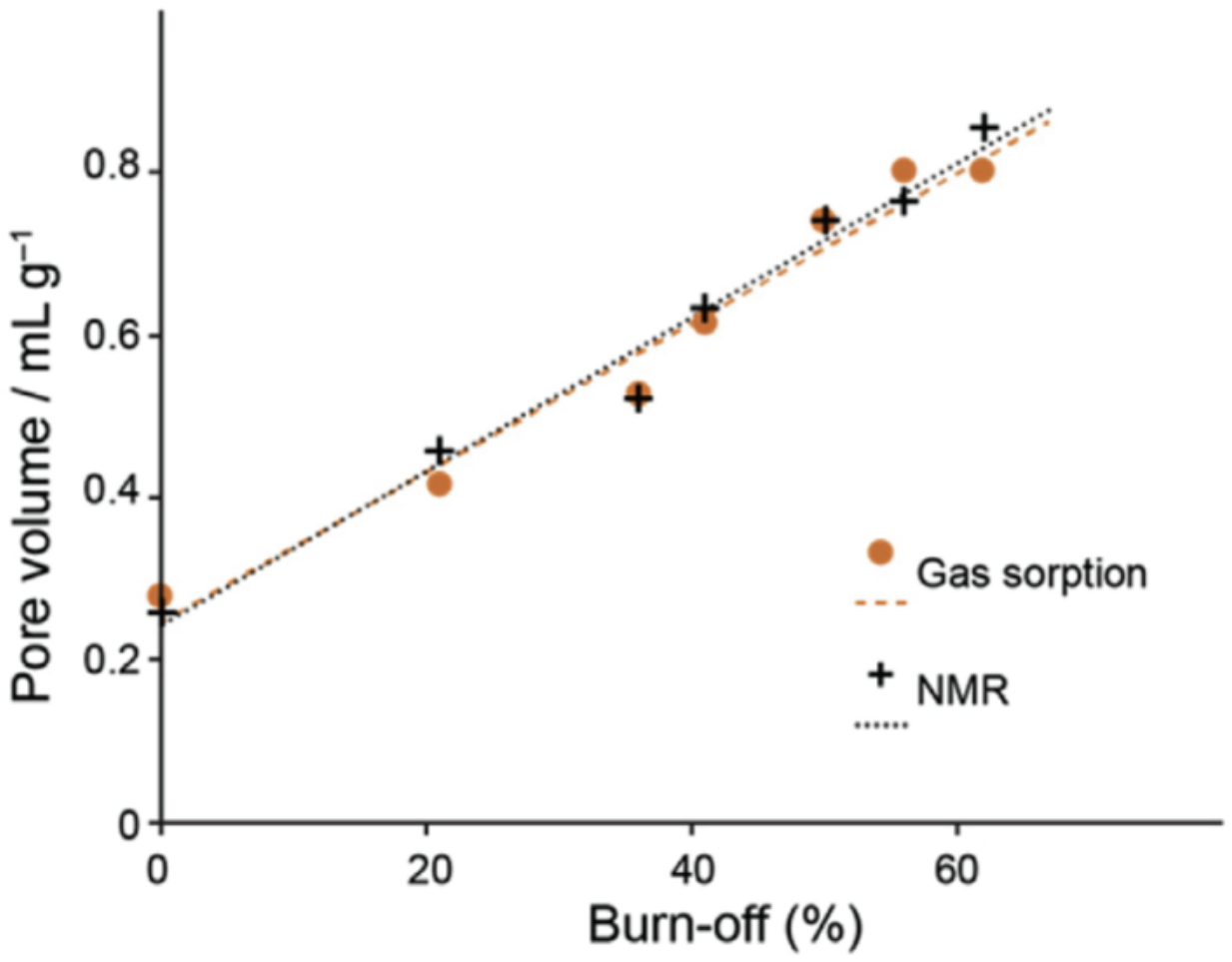

**Figure 16.** Comparison of pore volumes determined by conventional nitrogen gas sorption and an NMR method where $^{1}$H NMR experiments were used to quantify water adsorption. A series of PDCs were studied with different burn-off levels. Reproduced with permission from Ref. 46.

Beyond measuring pore volumes at saturation, NMR experiments on carbon materials may be applied more generally to measure gas adsorption isotherms. In this approach NMR experiments are obtained at a series of different gas pressures at constant temperature. Similar to traditional volumetric or gravimetric adsorption measurements, "NMR adsorption isotherms" also allow the determination of the gas adsorption isotherm through quantitative measurements. The advantage of the NMR approach is that additional information can be obtained from the spectra. This point is well illustrated by a $^{1}$H NMR study of water vapour adsorption in a PDC where NMR spectra enable the nucleation and growth of water clusters to be studied.[69] In an earlier study, Anderson *et al.* used a similar method to quantify $H_2$ adsorption isotherms in a series of porous carbon materials.[37] Key advantages of the NMR approach over traditional adsorption techniques are that both gases and liquids can be studied and local structural information can be obtained. NMR measurements do however have the limitation that the studied molecules must contain NMR active nuclei.

*5.4 Studying Carbon Structures Directly with NMR*

In principle, $^{13}C$ solid-state NMR could provide structural information about porous carbons and insight into *e.g.*, any defects present. However, studying disordered carbons by $^{13}C$ NMR is challenging owing to the extensive disorder and magnetic susceptibility effects that are present.[70, 71] Freitas & coworkers have shown that $^{13}C$ MAS NMR spectra of pyrolised chars generally show a broad resonance centred between ~ 110 - 130 ppm, consistent with the chemical shift expected for $sp^2$-hybridised carbons in graphene-like environments.[22, 72-75] The linewidth is often broader than the range of chemical environments expected for the samples (up to 200 ppm), suggesting that magnetic susceptibility broadening provides a significant contribution and therefore that linewidths cannot be directly related to the structural disorder present. For samples containing significant amounts of defects such as hydrogen- or oxygen-containing functional groups, distinct chemical environments can be observed;[23, 76] however, such systems are chemically quite dissimilar to those typically used for adsorption applications.

One way in which $^{13}C$ NMR has been used indirectly for the study of adsorption and diffusion is in reverse cross-polarisation (CP) experiments, where magnetisation is transferred from a $^{13}C$-enriched porous carbon surface to the nearby adsorbate species. Forse *et al.* have demonstrated that this is possible for $NEt_4$-$BF_4$ / $CD_3CN$ adsorbed on $^{13}C$-enriched TiC-CDC-1000, where in a static CP experiment, the selective $^{13}C \rightarrow {}^{1}H$ CP transfer was used to filter the poorly-resolved static spectrum so that only the in-pore $NEt_4$ cations were observed.[50] Subsequent work developed this approach further for the study of a 1-Methyl-1-propylpyrrolidinium bis(trifluoromethanesulfonyl)imide ($Pyr_{13}$TFSI) ionic liquid electrolyte via $^{13}C \rightarrow {}^{1}H$ and $^{13}C \rightarrow {}^{19}F$ CP for the respective observation of $Pyr_{13}$ cations and TFSI anions adsorbed on $^{13}C$-enriched TiC-CDC-1000.[77] In this system, the in-pore ions are poorly resolved even under MAS conditions due to the high viscosity and reduced diffusivity of the electrolyte, therefore the selective nature of the CP experiment is useful for distinguishing them from the ex-pore electrolyte. The CP kinetics were also found to be relatively slow, showing that although confined within the pores, the in-pore ions remain relatively dynamic. One practical challenge with the CP approach is that the conductive nature of the porous carbon coupled with the presence of highly ionic electrolyte solutions can lead to excessive sample heating, particularly when long CP pulses are used; however, this problem can be mitigated by active sample cooling. Further development of the reverse CP approach could facilitate observation and quantification of ions in systems where resolution is limited, and also provide information on dynamics through measurement of the CP kinetics.

## 6. Quantifying Dynamics and Diffusion of Adsorbed Species

*6.1 Effects of In-pore Exchange on the Ring Current Shift.*

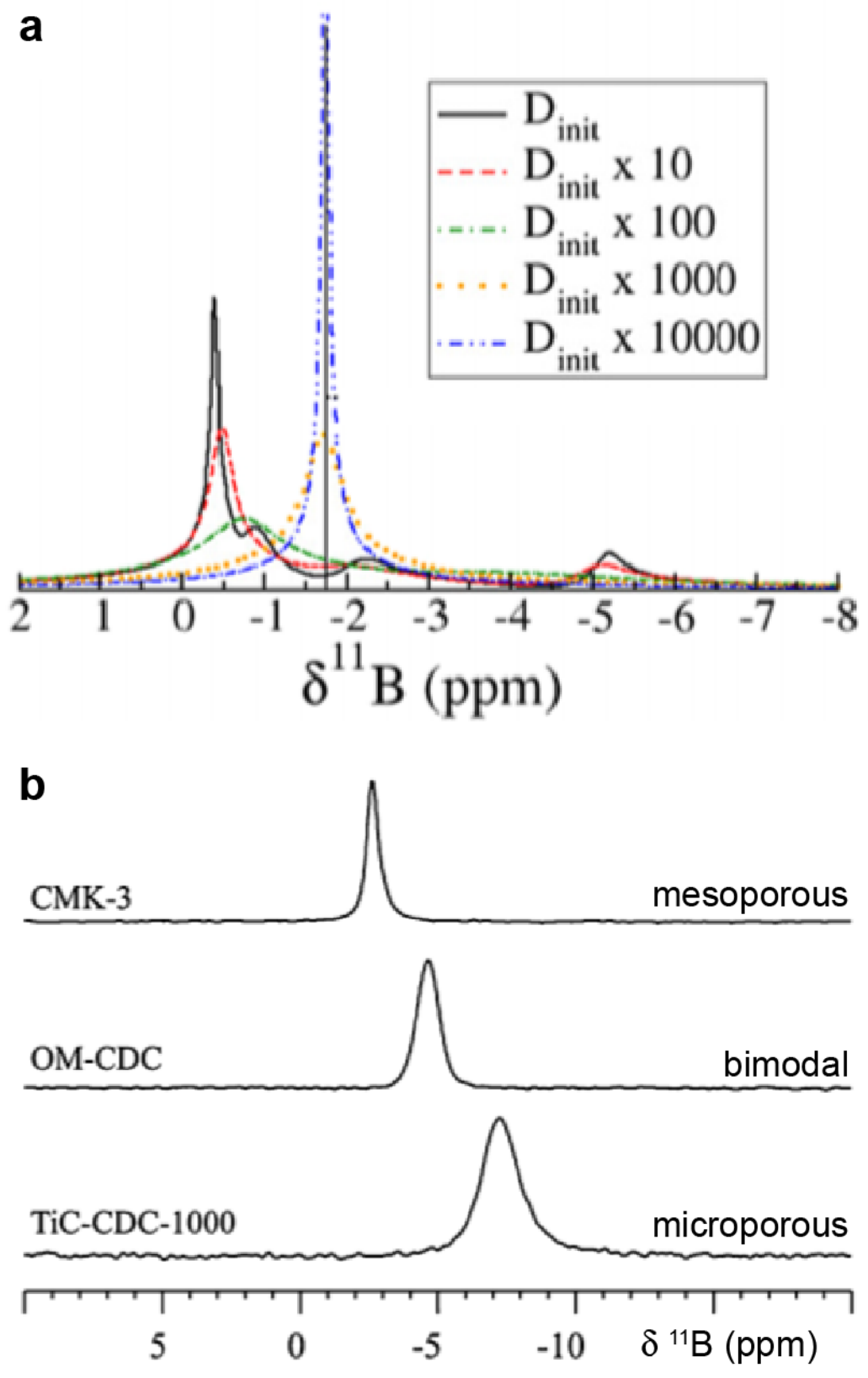

**Figure 17.** (a) Variation of the simulated in-pore lineshape with diffusion coefficient for species adsorbed within a single slit pore. As the diffusion coefficient is increased, peaks for different adsorption layers merge due to exchange averaging. [Reproduced with permission from Ref. 78]. (b) $^{11}B$ MAS NMR spectra of $NEt_4$-$BF_4$ / ACN adsorbed on carbons containing mesopores (CMK-3), micropores (TiC-CDC-1000) and both types of pore (OM-CDC) [Reproduced with permission from Ref. 45].

As well as providing information about the carbon pore structure, NMR measurements on adsorbed species can be used to probe their dynamics within the porous structure. Indeed, as discussed in the previous sections, most studies which have attempted to correlate the ring current shift with the pore width have accounted for averaging due to fast dynamics within the pore under consideration.[17, 46, 47] In general, this is found to give better agreement with experimental results, suggesting that adsorbed species are highly mobile within the pores and explore the width of the pore on a timescale which is must faster than the shift difference between different locations within the pore. The effects of intra-pore dynamics were confirmed in a theoretical study by Merlet *et al.*, who employed a lattice simulation method to explore the effect of exchange between different adsorption layers in a 4 nm model slit pore.[78] In the ultra-slow exchange regime, in-pore resonances corresponding to distinct adsorption layers are observed across a range of –0.5 to –5 ppm, which is not typically observed experimentally. As exchange between the layers was introduced into the simulation (also accounting for the free energy profile of the slit pore), the in-pore resonances merged into a single resonance at a weighted average shift of –2 ppm (Figure 17a). This is consistent with the observation of a single in-pore resonance for most porous carbons that have been studied.

As well as being sensitive to dynamics within a single pore, ring current shifts measured for adsorbed species have revealed dynamics *between* connected pores within the carbon structure. This was illustrated by Bordchardt *et al.*, who studied the adsorption of $NEt_4$-$BF_4$ / ACN on OM-CDC, which has a characteristic bimodal pore size distribution with pores of 1.0 and 4.1 nm.[45] Based on the pore-width dependence of the ring current shift, the large variation in pore width within OM-CDC should result in two separate in-pore resonances; however, only a single resonance was observed with Δδ of –3.6 ppm (Figure 17b). This shift is intermediate between that of the same electrolyte adsorbed on TiC-CDC-1000 (pore size 1.0 nm; Δδ = –6.2 ppm) and CMK-3 (pore size 4.5 nm; Δδ = –1.7 ppm), suggesting that exchange between the 1.0 and 4.1 nm pores in OM-CDC results in averaging of the ring current shifts for both pore environments. Based on the 4.5 ppm difference in Δδ for TiC-CDC-1000 and CMK-3, the averaged shift observed for OM-CDC suggests that exchange between the micropores and mesopores takes place on the millisecond timescale. As discussed in Section 5, similar observations have been made in NMR studies of other porous carbons, where a single in-pore resonance has been observed for carbons with multi-modal or hierarchical pore size distributions as measured by gas sorption analysis.[46, 47] However, despite the inter-pore exchange averaging that often takes place, for many systems the in-pore resonance is still broader than the ex-pore resonance and varies significantly with temperature,[44] suggesting that the interpore dynamics are not fully in the fast exchange regime, and/or that a distribution of locally-averaged shifts remains.

The factors contributing to the in-pore linewidth were investigated by Alam & Osborn Popp, who used 2D exchange experiments to separate homogenous and inhomogenous contributions.[51] For a carbonate solvent adsorbed on a commercial microporous carbon, in a short mixing-time (10 μs) spectrum the in-pore resonance was concentrated along the diagonal line showing that the observed linewidth was dominated by inhomogeneous broadening, *i.e.*, a distribution of in-pore shifts arising from incomplete exchange averaging between different pore environments. The homogenously-broadened linewidth was found to be approximately half of the inhomogenous component, and was attributed to either residual dynamic effects or incomplete averaging of the anisotropic magnetic susceptibility (which is not averaged by MAS[49]). In a long mixing-time (100 ms) spectrum, the in-pore resonance became equally broadened in the on- and off-diagonal dimensions, showing that the adsorbed solvent molecules experience the full chemical shift distribution (*i.e.*, explore all pore environments) on this timescale. However, while these observations are qualitatively consistent with interpore exchange, relating the in-pore linewidth to either the pore structure or dynamic timescale in the absence of additional information remains a challenge. Merlet *et al.* extended the lattice simulation method to gain further insight by exploring the effects of both the pore connectivity and experimental conditions on the observed lineshape.[78] By comparing a random distribution of pore widths with a model assuming a gradual variation, or gradient, of pore widths across the carbon particle, it was found that the spatial connectivity of pores can influence the in-pore lineshape. However, by reducing the energy barrier associated with the diffusion of the adsorbed species in the gradient model, a similar lineshape to the random model was obtained. This shows that it is difficult to interpret the spatial distribution of the pores on the sole basis of the lineshape of the in-pore resonance without some additional knowledge of the adsorbent-adsorbate interactions. The exchange-averaged lineshape was found to be further influenced by the temperature and Larmor frequency, which affect the rate of exchange between pores of different widths and the differences in absolute frequency associated with the ring current shifts, respectively. Notably, the temperature and Larmor frequency dependences showed marked differences for the different spatial arrangements of pores considered. This information could be used together with experimental data to distinguish between different pore arrangements in porous carbon materials.

*6.2 Effects of In-pore - Ex-pore Exchange.*

When an external reservoir is in contact with the carbon particles (*i.e.*, when the sample is saturated with excess adsorbate), adsorbed species can undergo exchange between the in-pore and ex-pore environments. This was qualitatively illustrated by Wang *et al.* who observed cross peaks at a mixing time of 500 ms in a $^{11}$B static 2D exchange spectrum of $NEt_4$-$BF_4$ / ACN adsorbed on YP17 activated carbon, suggesting that mixing between the in-pore and ex-pore environments takes place approximately two orders of magnitude slower than the millisecond timescale inter-pore exchange process.[79] To more accurately quantify the in-pore – ex-pore exchange timescale, Griffin *et al.* performed a systematic study on the same electrolyte-carbon system, whereby a set of 2D exchange spectra were recorded with a range of mixing times.[80] In principle, the rate constant for a two-site exchange process can be determined from a simple fit of the build-up of the integrated cross-peak/diagonal peak intensity ratios as a function of mixing time.[81] Using this approach, the exchange timescale was again found to be on the order of hundreds of miliseconds. However, the experimental build-up of the intensity ratio did not fit well to the two-site model, suggesting that the exchange behaviour is more complex than the simple two-site exchange model assumes. The exchange kinetics of $NEt_4$-$BF_4$ / ACN adsorbed on YP50 were subsequently examined in more detail by Fulik *et al.*, who also found poor agreement with the two-site exchange model but obtained a significantly improved fit when two separate exchange processes were assumed.[43] The fits revealed a "slow" component on the timescale of seconds, as well as a "fast" component on the millisecond timescale. The slow exchange process was attributed to the diffusion of species from within the interior volume of the carbon particles to the free liquid. These species are required to diffuse through the carbon pore network before in-pore – ex-pore exchange can take place at the carbon surface. Although exchange between connected pores can take place on the millisecond timescale (as discussed above), macroscopic diffusion of adsorbed species from the interior regions of carbon particles to the exterior surface is restricted by the tortuousity of the pore network and therefore occurs on a much slower timescale. The fast process was attributed to in-pore – ex-pore exchange taking

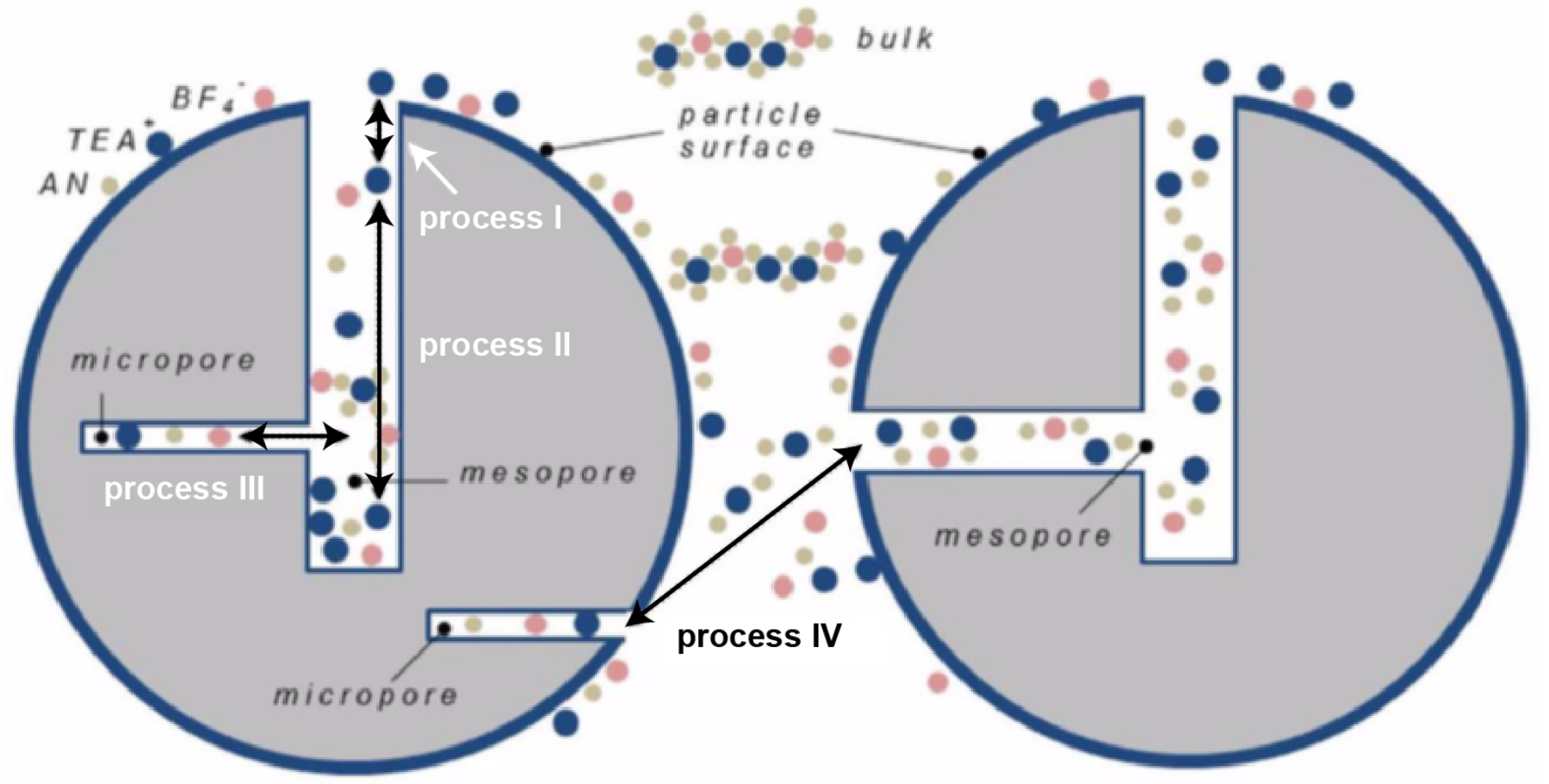

**Figure 18.** Exchange processes proposed by Fulik *et al.*[43] Process I: Exchange within a thin surface layer, i.e., between pores close to the surface and the external bulk surface layer at the carbon particles. Process II: Diffusion from the interior of the particle to the surface. Process III: Exchange between connected pores. Process IV: Exchange between different particles. [Reproduced with permission from Ref. 43]

place close to exterior surfaces of carbon particles. In the interfacial region between the carbon pore network and the free liquid, exchange can take place very rapidly because adsorbed species have a relatively small distance to travel to move between the two environments. Further processes were identified corresponding to exchange between micropores and inter-particle exchange via free liquid (Figure 18). It should be noted that this model remains a simplification of the real exchange kinetics, which in principle should be described by a continuum of processes reflecting the continuous variation in exchange kinetics between the ex-pore environment and progressively deeper locations within carbon particles. However, the division into separate local processes remains a useful concept for differentiating timescales of the diffusion processes present in the system.

The large difference in timescale between the bulk and interfacial diffusion processes suggests that the overall in-pore – ex-pore exchange kinetics should be influenced by the carbon particle size. For a given electrolyte-carbon system, smaller particles should comprise a greater proportion of the interfacial region with fast in-pore – ex-pore exchange kinetics. This hypothesis was investigated by Cervini *et al.*, who performed exchange experiments on two samples of microporous PDCs with average particle sizes of 80 and 21 μm.[82] For the larger particle size, rate constants of 57 and 4 Hz were determined for the two processes, while for the smaller particles rate constants of 597 and 50 Hz were determined, reflecting an order of magnitude increase in overall in-pore – ex-pore exchange kinetics in line with a prediction based on a simple spherical particle model. In principle, the two-process model would predict that the rate constants for the two processes should remain constant, whereas their relative contributions to the exchange kinetics should change with particle size. The variation in the rate constants observed in this work probably reflects the limitations of the two-process model, which also becomes less valid for small particle sizes. In addition, the PDC samples studied showed appreciable particle size distributions ranging over several tens of microns, meaning that the measured exchange kinetics should be the superposition of range of kinetic parameters weighted by the particle size distribution.

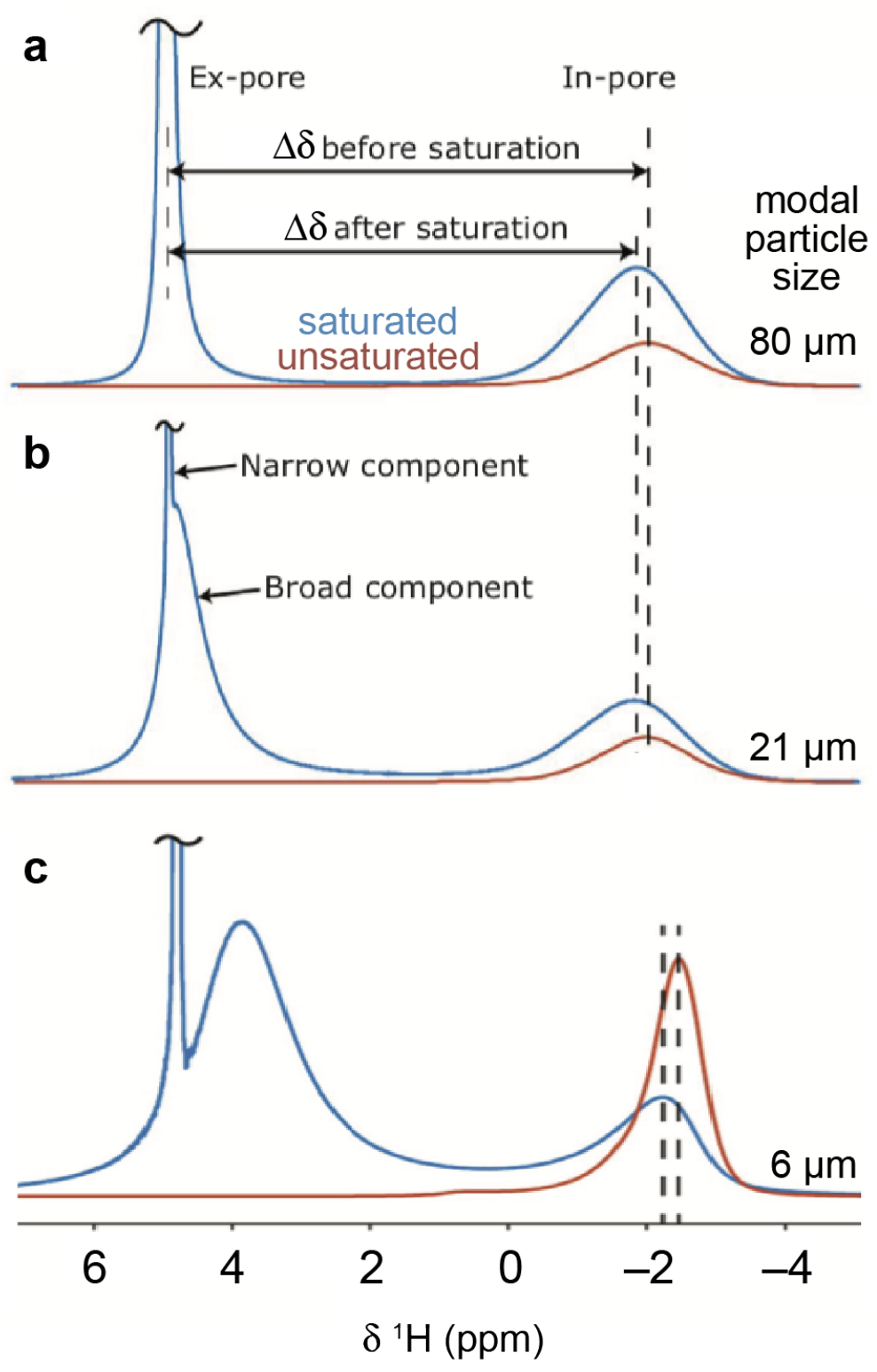

**Figure 19.** $^1$H MAS NMR spectra of $H_2O$ adsorbed on (a, b) PDCs and (c) YP50F with respective particle sizes of 80, 21 and 6 μm. Spectra for saturated and unsaturated samples are shown in blue and red, respectively. In-pore – ex-pore exchange in saturated samples with smaller particles leads to partial exchange averaging of the in-pore and ex-pore resonances. [Reproduced with permission from Johnson Matthey Technol. Rev., 2020, 64, (2), 152. https://www.technology.matthey.com/article/64/2/152-164/

Another manifestation of in-pore – ex-pore exchange is partial averaging of the ring current shift. If the in-pore – ex-pore exchange rate is comparable to the absolute frequency of the ring current shift, partial averaging of the in-pore and ex-pore peaks can be observed. This was illustrated by Fulik *et al.* who compared NMR spectra of $NEt_4$-$BF_4$ / ACN adsorbed on YP50F for unsaturated and saturated samples.[43] Upon saturation, the onset of in-pore – ex-pore exchange between the adsorbed species and free liquid resulted in shifting and asymmetric broadening of the adsorbate resonances. This effect was particularly marked for the $^2$H NMR resonance of the fast-exchanging deuterated ACN solvent molecules, for which the in-pore shift increased by 1.2 ppm. The ionic electrolyte species showed similar effects but the in-pore shift change was much less pronounced, reflecting their reduced mobility relative to the solvent molecules. Cervini *et al.* showed that exchange averaging also shifts and broadens the ex-pore resonance, and that particle size is also an important factor.[82] For 80 μm PDC particles saturated with $H_2O$, negligible exchange averaging was observed meaning that the measured Δδ value was a relatively accurate representation of the ring current shift (Figure 19a). However for 21 μm particles, the ex-pore resonance was shifted and broadened relative to neat $H_2O$ (Figure 19b). The same work also compared a $H_2O$-saturated sample of YP50F with a much smaller average particle size of 6 μm with a particle size distribution of approximately 15 μm. For this system chemical shift of the ex-pore $^1$H resonance decreased by almost 1 ppm while the in-pore resonance increased by 0.7 ppm upon saturation. Significant broadening of both resonances was also observed (Figure 19c). These observations highlight that the partial averaging effects of in-pore – ex-pore exchange must be taken into account when measuring Δδ. If significant exchange is present, the measured difference between the free liquid/ex-pore and the in-pore chemical shifts may not be representative of the true ring current shift. This phenomenon is particularly pronounced for carbons with small particles sizes or for highly mobile adsorbates where the exchange rate can be comparable to the absolute in-pore – ex-pore frequency difference at standard magnetic field strengths. As Fulik *et al.* have pointed out,[43] accurate measurements of Δδ should only be made on unsaturated samples where exchange effects are inhibited due to the absence of the free liquid reservoir.

### *6.3 Quantifying Diffusion in Porous Carbons*

Pulsed-field gradient (PFG) NMR experiments offer a direct measurement of the self-diffusion coefficients of molecules adsorbed in porous carbons. In these experiments the effective self-diffusion coefficient, $D$, is determined during a period of time, $\Delta$, known as the observation time (or diffusion time).[83, 84] An early study probed the diffusivities of small molecules that were loaded into a series of activated carbons by vapour dosing.[85] In this extensive study it was shown that the self-diffusion of both adsorbed water and ethanol increased as a function of the average carbon pore size (Figure 20). Notably, the self-diffusion of ethanol was found to generally be slower than that of water for a given carbon. The authors rationalised this observation by considering the increased steric hindrance experienced by the larger ethanol molecules as they move through the carbon pore network. In a further study, PFG NMR measurements were made for carbons loaded with mixtures of adsorbates,[86] highlighting the potential of PFG NMR to study complex multi-component adsorbent-adsorbate systems.

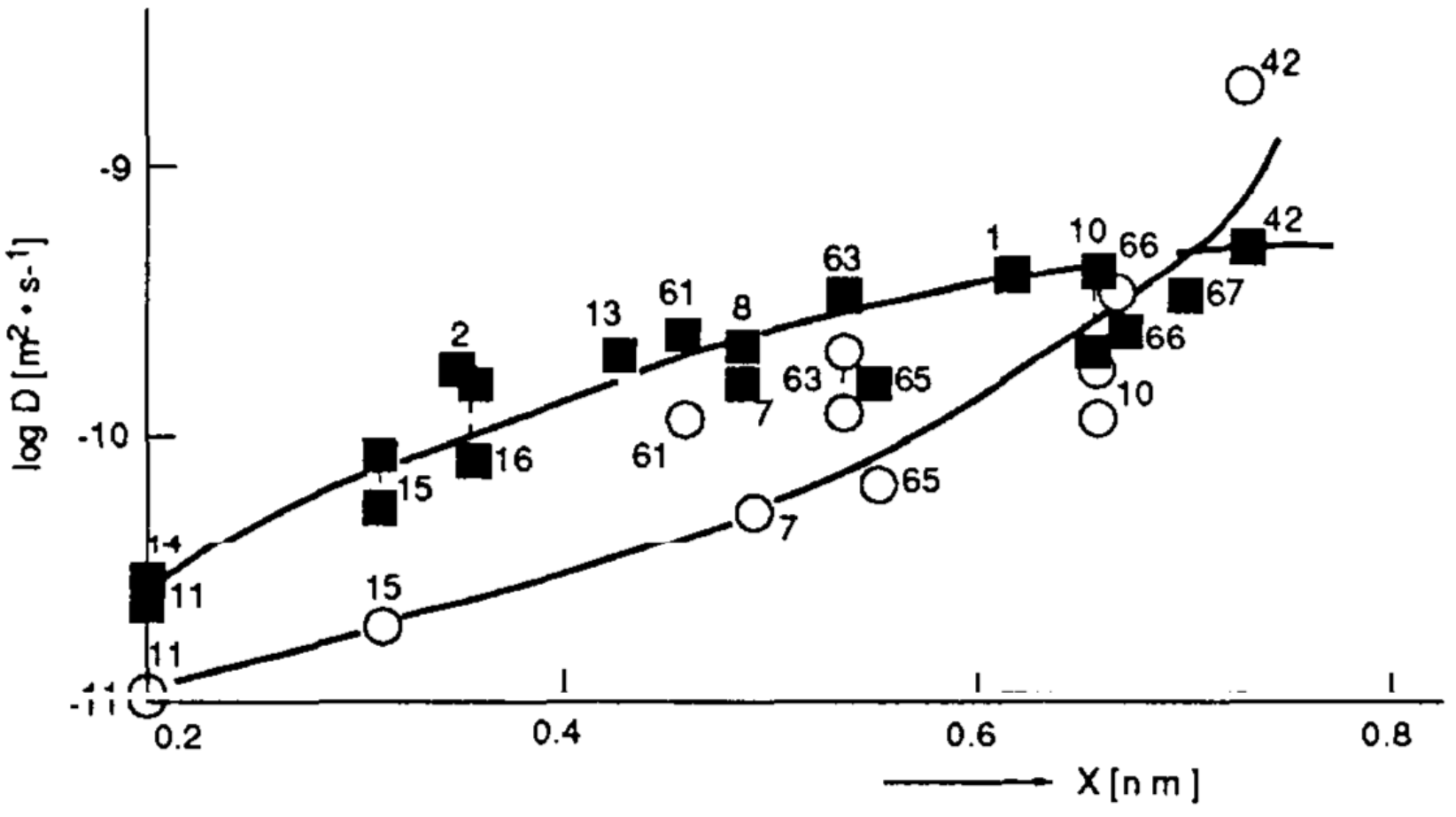

**Figure 20.** Self-diffusion of adsorbed water (filled squares) and adsorbed ethanol (open circles) in a series of different carbon materials with different mean half width pore sizes, *X*. [Reproduced with permission from Ref. 85]

For experiments with vapour dosing, as above, it can be assumed that the adsorbate resides solely in the carbon micro/mesopores, with no significant amount of ex-pore adsorbate. This makes it relatively straightforward to measure in-pore self-diffusion. However, for studies of carbons saturated with liquids, both in-pore and ex-pore adsorbate species are anticipated. The ring current chemical shift for in-pore species then provides a very convenient means to separately study the diffusion of in-pore and ex-pore molecules, via PFG NMR experiments. Figure 21 exemplifies this point and shows pulsed-field gradient NMR spectra for electrolyte ions in the working electrode of a supercapacitor cell with no applied voltage. A series of stimulated echo experiments with bipolar-pulsed field gradients are shown (Figure 21a), in which the pulsed field gradient gets successively stronger. The rapid decay of the ex-pore / free electrolyte cation signal as a function of the pulsed-field gradient strength is evident in Figure 21a, correlating with the rapid self-diffusion of these species. In contrast, a much slower signal decay is observed as a function of the gradient for the in-pore ions, due to their slower self-diffusion. Analysis of the in-pore intensities as a function of the "*b*-value" (which depends on the square of the gradient strength and a number of experimental constants) showed that a biexponential function was required to obtain a good fit (Figure 21b). The fast diffusing component was assigned to in-pore ions that have undergone in-pore – ex-pore exchange processes during the observation time, $\Delta$, while the slow diffusing component was assigned to in-pore ions that did not participate in in-pore – ex-pore exchange during the observation time. This assignment was supported by a series of PFG NMR measurements with different observation times, which found that the exchange component was almost absent in experiments with small observation times (since there was insufficient time for exchange), while the exchange component dominated for large observation times (all ions undergo in-pore – ex-pore exchange at long times, see also Figure 21b below). Similar measurements and analysis were performed for the electrolyte *anions* via $^{19}F$ PFG NMR measurements (Figure 21c,d). The PFG NMR measurements thus enable studies of both self-diffusion and in-pore – ex-pore exchange processes.

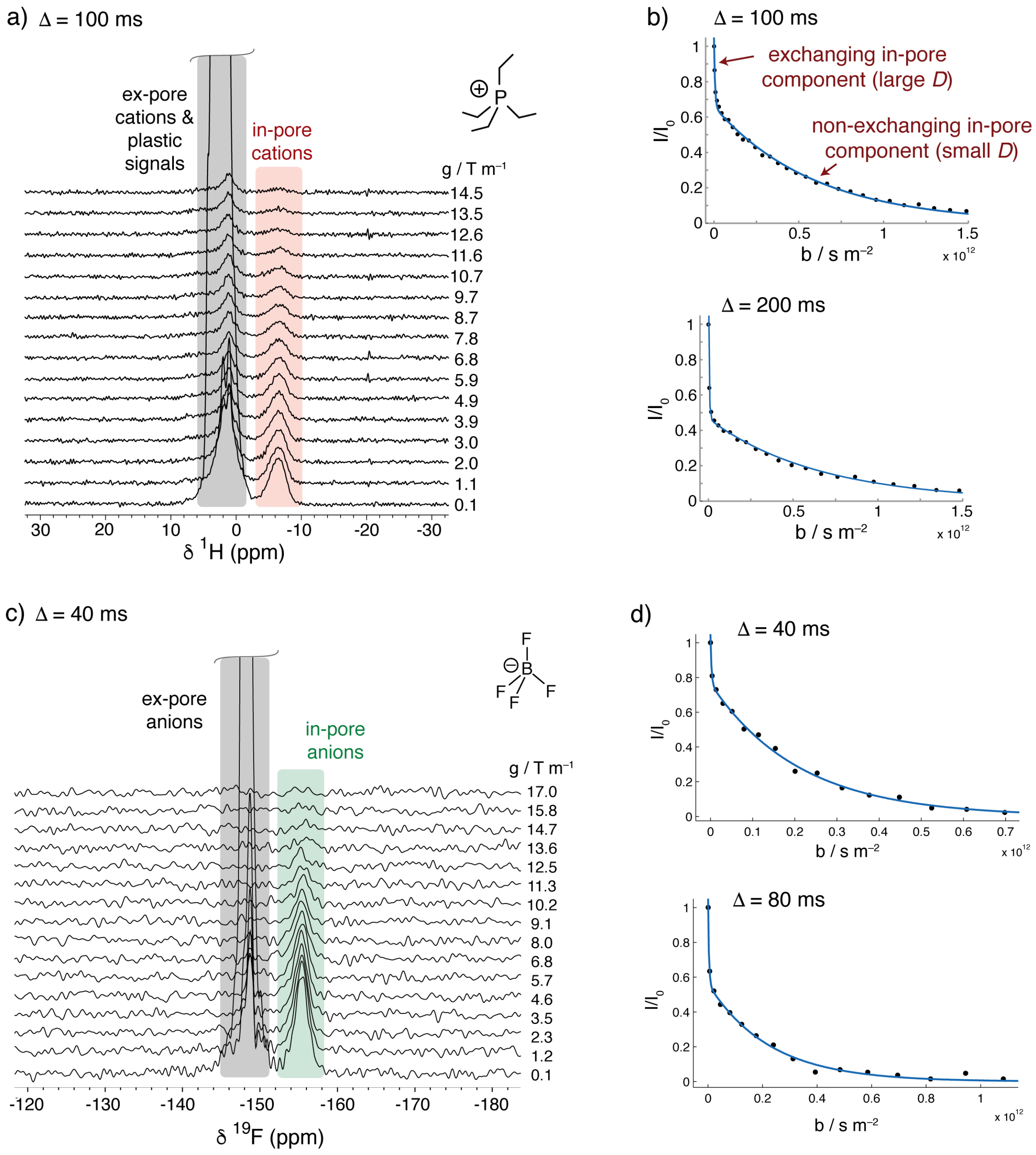

**Figure 21.** PFG NMR spectra and analysis on the working electrode (YP50F porous carbon) of a supercapacitor cell with no applied voltage (i.e. 0 V). a) and c) show $^{1}H$ and $^{19}F$ NMR spectra that probe the cations and anions, respectively, of the $PEt_4BF_4$ (1.5 M) in deuterated acetonitrile electrolyte. b) and d) show the integrated intensity of the in-pore resonance (normalised by the intensity for the experiment with the weakest gradients), plotted against the *b* value, which depends on the square of the gradient strength and a number of experimental constants. Additional experimental details are available in Ref. [87]. [Reproduced with permission from Ref [87]].

The key findings of this PFG NMR study are summarised in Figure 22.[87] In the absence of an applied potential it was found that in-pore self-diffusion coefficients (*i.e.* the non-exchanging in-pore species) in YP50F carbon electrodes were at least two orders of magnitude smaller than the neat electrolyte diffusion coefficients, for both anions and cations (Figure 22a), highlighting the dramatic impact of adsorption on molecular diffusion. A second activated carbon with a larger average pore size, YP80F, revealed faster in-pore self-diffusion, showing that carbon pore size plays an important role, similar to the findings of the early work on water and ethanol diffusion described above.[87] Figure 22b shows the fraction of in-pore species *not* undergoing exchange during the observation time, denoted $A_{\text{in-pore}}$. The faster the decay of this curve, the faster the in-pore – ex-pore exchange. Figures 22a and 22b show the connection between in-pore diffusion and exchange – the faster the diffusion of the in-pore species, the more rapid in-pore – ex-pore exchange processes. This is consistent with the aforementioned work by Fulik *et al.*[43] who treated the overall in-pore – ex-pore exchange kinetics as being influenced by in-pore diffusion processes within the particles. Mean lifetimes for the anions and cations inside the carbon particles were extracted through exponential fits of the curves in Figure 22b.

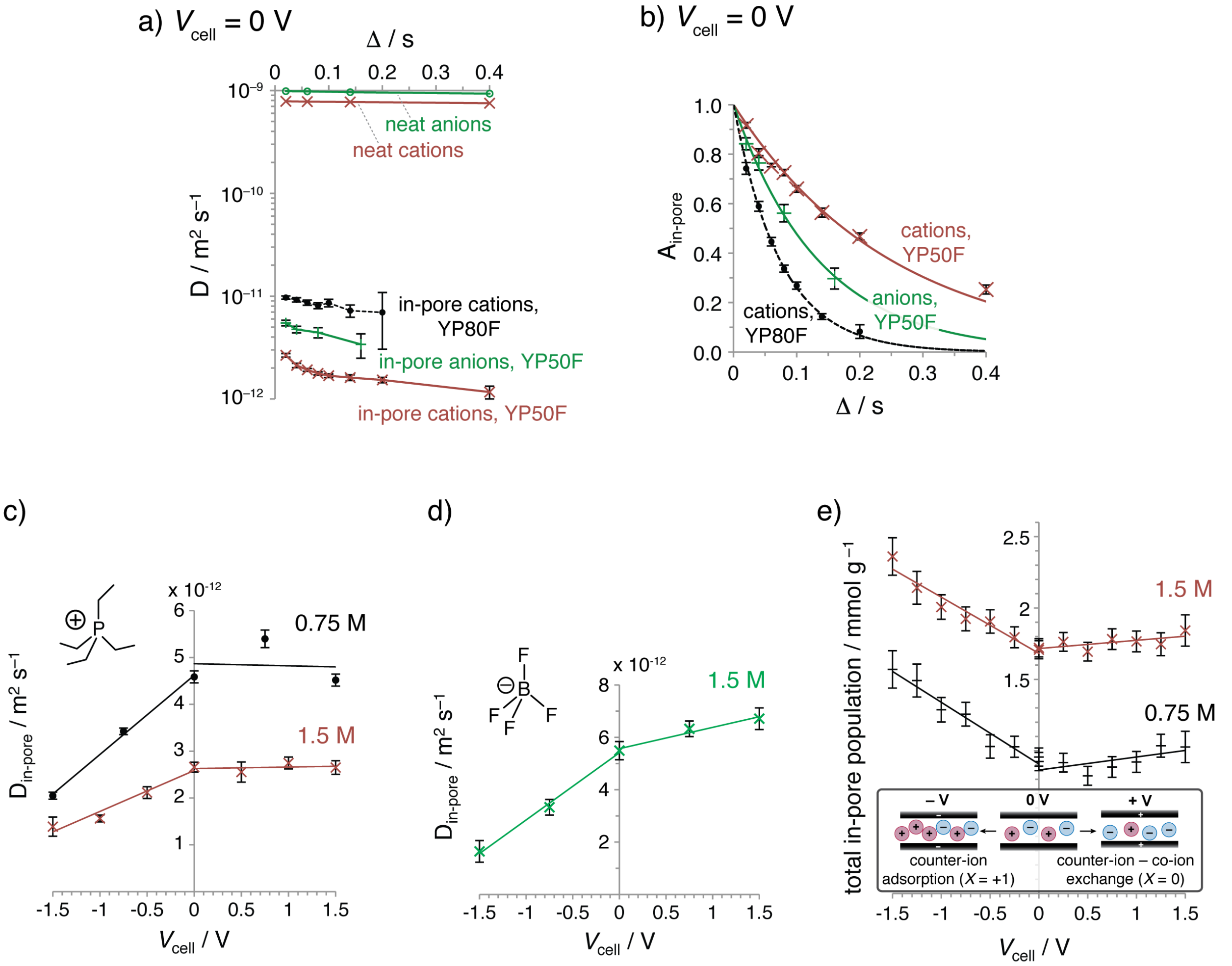

**Figure 22.** Key results of PFG NMR experiments performed on supercapacitors with porous carbon electrodes and $PEt_4BF_4$ in deuterated acetonitrile electrolyte. a) Self-diffusion coefficients determine for electrolyte anions and cations in neat electrolyte and in the pores of carbon electrodes (1.5 M $PEt_4BF_4$ in deuterated acetonitrile electrolyte). Data are shown for two carbons, YP50F and YP80F (YP80F has a larger average pore size). b) Fraction of non-exchanging in-pore ions, $A_{\text{in-pore}}$, as a function of the observation time, $\Delta$, in the PFG NMR experiment. The decay of this data reflects the increasing amount of in-pore – ex-pore exchange at larger observation times. c), d) In-pore self-diffusion coefficients

for anions and cations, respectively, as a function of charging potential. e) Total in-pore ion populations at the cell voltages explored in c) and d). Reproduced with permission from Ref. [87]

Lastly, experiments performed on supercapacitors at different charging potentials revealed charging-dependent self-diffusion behaviour (Figure 22c, d). When the electrode under investigation was positively charged (positive $V_{cell}$), only very minor changes in the in-pore self-diffusion coefficients were observed for both cations and anions (Figure 22c,d). In contrast, charging negatively (negative $V_{cell}$) brought about significant reductions in the in-pore self-diffusion coefficients. This effect was rationalised by considering the charge storage mechanism of this system (Figure 22e, discussed in more detail in Section 7). Upon positive charging an ion exchange mechanism operates,[88] leading to a fairly constant total population of ions as a function of voltage. In contrast, upon negative charging a counter ion adsorption charging mechanism leads to an increase in the total in-pore ion population. It was hypothesised that this denser packing of ions leads to a decrease in their self diffusion coefficients, which was corroborated by additional measurements with lower electrolyte concentrations Figure 22c.[87]

Additional studies have further explored the impact of the carbon pore structure on diffusion in activated carbons. One study challenged the commonly held notion that porous carbons with hierarchical pore structures should yield faster adsorbate self-diffusion.[89] In-pore self diffusion coefficients for the anions, cations and solvent of the supercapacitor electrolyte $NEt_4BF_4$ ($TEABF_4$) (1 M) in deuterated acetonitrile (ACN) were measured in three materials, a microporous carbon, a mesoporous carbon, and a hierarchical carbon with both mesopores and micropores (Figure 23). Large reductions of in-pore diffusion (compared to bulk electrolyte) were observed for all species in the microporous carbon, with $D_{bulk}/D_{confined}$ approaching 1,000 (Figure 23), similar to the findings in Figure 22c. In contrast in the mesoporous carbon, much smaller $D_{bulk}/D_{confined}$ values on the order of 10 were measured, reflecting the much more rapid diffusion in mesopores. Surprisingly, in the hierarchical carbon the confined cation and anion self-diffusion coefficients were similar to those observed for microporous carbon, while the solvent diffusion compared more closely to the mesoporous carbon. The authors rationalised this unexpected effect by suggesting that preferential ion adsorption in the micropores of the hierarchical carbon gives rise to the measured diffusion behaviour. Further studies should be carried out to explore the effect of carbon pore size and structural heterogeneity on self-diffusion in carbonaceous materials.

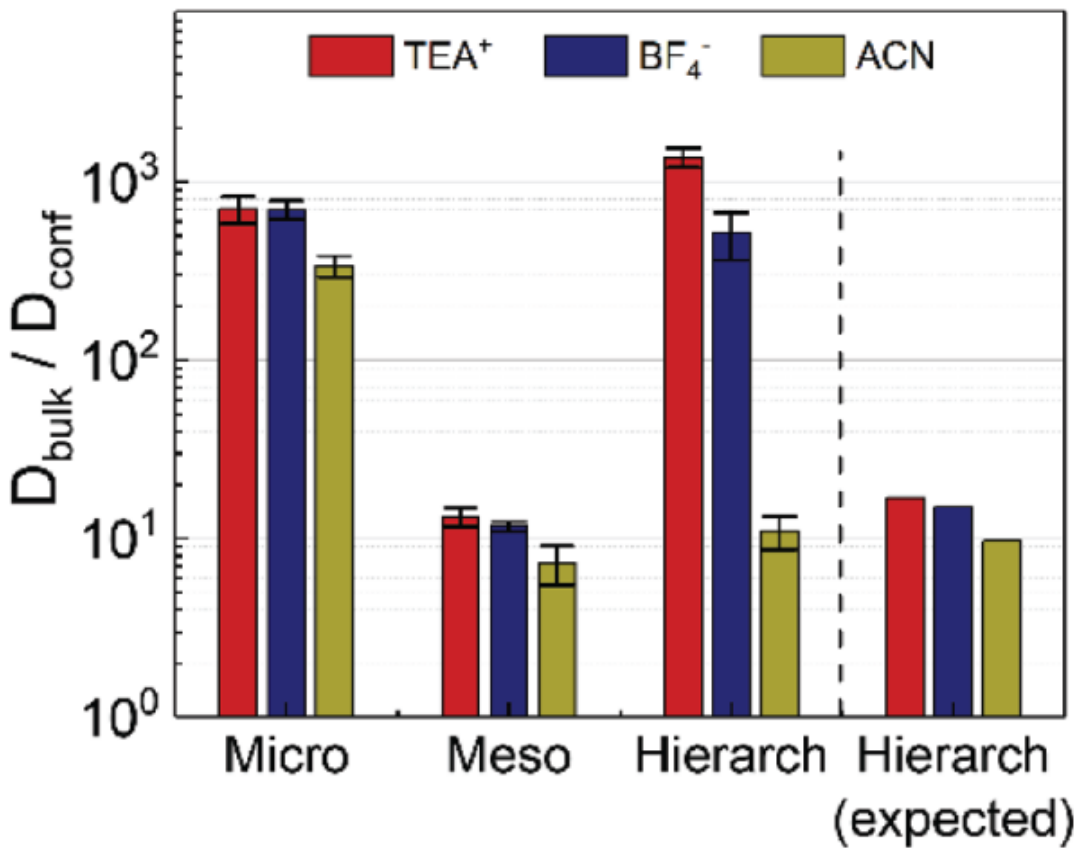

**Figure 23.** Effect of carbon porosity on self-diffusion of tetraethylammonium (TEA) cations, tetrafluoroborate ($BF_4$) anions and acetonitrile (ACN) solvent in carbons with different types of porosity. Diffusion data are shown as $D_{bulk}/D_{confined}$, where $D_{bulk}$ is the self-diffusion for bulk electrolyte, and $D_{confined}$ is the self-diffusion for pore-confined species in the carbons under study. Reproduced with permission from Ref. [89].

## 7. Studying Electrosorption of Ions in Porous Carbons by NMR Spectroscopy

### *7.1 Principles of Electrosorption*

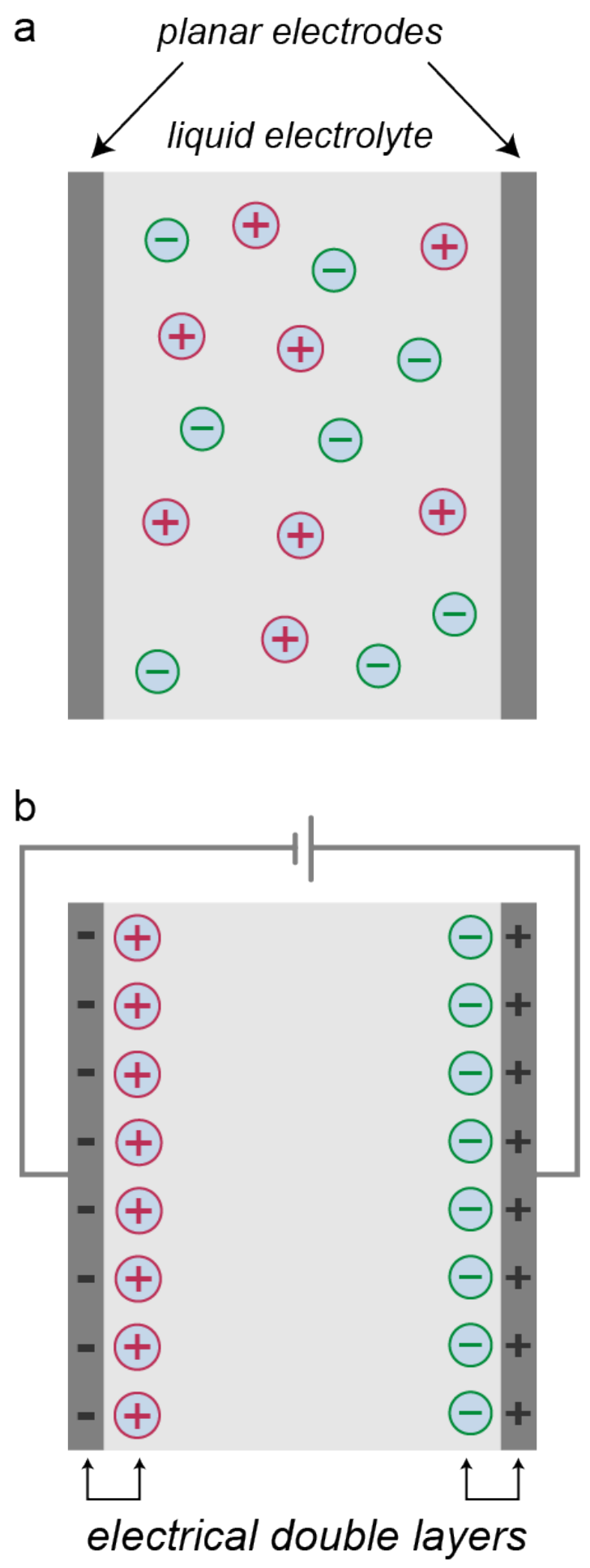

Figure 24. Illustration of the simplified mechanism of electrosorption of liquid electrolyte ions on the surfaces of charged planar electrodes. (a) Ions are initially in free solution. (b) Upon application of an external potential, electronic charge accumulates in the conducting electrodes, and this is balanced by an equal and opposite ionic charge within the electrolyte to form an electrical double layer.

In addition to spontaneous adsorption, ions within liquid electrolytes can be electrostatically adsorbed, or *electrosorbed*, on porous carbons by applying a voltage to the carbon surface. The applied voltage causes electronic charge to accumulate in the carbon surface, which is balanced at the interface by an equal and opposite charge associated with the electrosorbed ions from the electrolyte. Electrosorption is exploited for charge and ion storage in a number of applications, notably supercapacitors and capacitive desalination. For planar electrodes, electrosorption can be relatively well described in terms of classical electrical double-layer theory, where counter-ions (*i.e.*, those of opposite polarity to the electrode surface) form layers on the electrode surfaces (Figure 24).[90, 91] However, for microporous carbon electrodes with pore sizes comparable to the solvated ion size, confinement effects mean that the same layered structures of ions close to the carbon surface cannot occur. Modelling studies have shown that the electrosorption mechanism is more complex, involving local rearrangements of ion positions as well as ion migration into and out of the pores.[78, 92] Despite this, many aspects of the electrosorption mechanism in porous carbons are still poorly understood and there are few experimental techniques that can probe the structure of the electrical double layer. The fact that NMR can distinguish species inside the pores from those in the bulk electrolyte makes it one of the few techniques that can quantify adsorbed ions, and track how their populations and local environments change during electrosorption.

### *7.2 Quantifying In-pore Ions in the Absence of an Applied Potential*

Before considering electrosorption, it is important to quantify the adsorption of electrolyte species at 0 V since this determines the starting point for the movement of ions into, or out of, the micropores during charging. As discussed in the previous sections, most NMR studies show that liquid adsorbates spontaneously fill carbon micropores. For organic electrolytes, the density of in-pore ions can be high. For 1.5 M $NEt_4$-$BF_4$ / ACN adsorbed on YP50F, $BF_4$ anions were adsorbed at a density of 0.86 mmol per gram of carbon.[80] The same density was also measured in a separate study for tetraethylphosphonium tetrafluoroborate ($PEt_4$−$BF_4$) / ACN adsorbed on the same carbon.[88] Using estimated solvated ion diameters, the total volume of the in-pore electrolyte ions was found to be significantly higher than the available pore volume as measured by gas sorption. This counter-intuitive result contrasts the good agreement seen between NMR- and gas sorption-derived pore volumes when $H_2O$ is used as the adsorbate (as discussed in Section 5.3). This discrepancy was attributed to the dense packing of electrolyte ions inside the micropores such that their solvation shells overlap. This means that the effective solvated ion diameter inside the micropores is smaller than that in free solution. Even higher ionic packing densities have been measured for ionic liquid electrolytes, where there are no additional solvent molecules to take up any of the pore volume. For pure EMI-TFSI and $Pyr_{13}$-TFSI ionic liquids adsorbed on YP50F, 1.8 and 1.6 mmol of ions per gram of carbon were measured.[44] When the ionic liquid was diluted with ACN solvent to approximately 1.2 M, the in-pore ion populations reduced as the solvent molecules displaced some of the ions.

NMR studies of aqueous electrolytes have revealed some differences in their spontaneous adsorption behaviour. Although the bare ion sizes can be very small, strong interactions with the $H_2O$ solvent molecules can result in a larger effective ion size. It was shown that for a PDC with a relatively small average pore width (0.58 nm) soaked with NaF (0.8 mol $kg^{-1}$ aq), the ions were unable to enter the carbon pores despite the presence of in-pore $H_2O$ molecules as observed by $^1H$ MAS NMR.[93] Through comparison with measurements on a PDC with larger average pore size, this was confirmed to be due to steric effects where the strongly solvated ions were physically too large to fit inside the small pores. Steric effects have also been implicated in a recent systematic study of aqueous electrolyte adsorption in PDCs, where a reduction of the measured $\Delta\delta$ for adsorbed cations was observed for low electrolyte concentrations.[46] It was proposed that at low concentrations, solvated ions preferentially reside in larger pores where there is less distortion or rearrangement of the solvation shell; this indirectly reduces the ring current shift due to the larger average ion-carbon distance. This idea is in line with a number of theoretical studies which predict significant energy barriers for the entry of hydrated ions into small micropores, particularly for strongly-solvated Li and Na cations.[94-96] In addition to the concentration effect on the $\Delta\delta$ value, the same work also found significant specific ion effects. Comparing 1 M LiCl (aq) and CsCl (aq) adsorbed on the same PDC, the $\Delta\delta$ value for the Cs cations was 5 ppm larger. The reason for this difference is not yet fully understood but it was proposed that the smaller and weaker hydration shell of $Cs^+$ compared to $Li^+$ means that $Cs^+$ can more closely approach the carbon surface and is more prone to partial desolvation, which can have a significant impact on chemical shifts of ions.[93, 97]

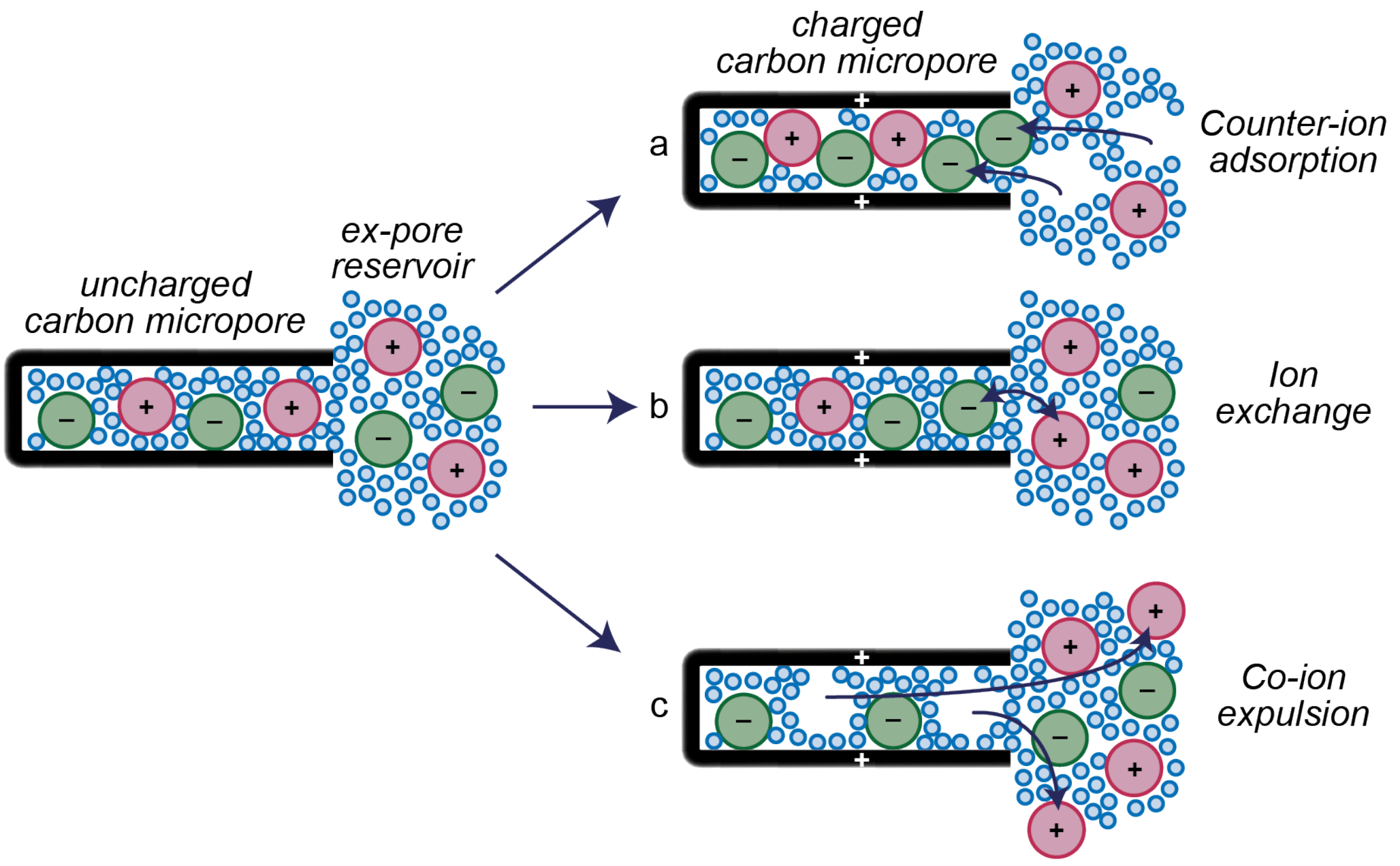

Figure 25. Schematic illustrations of possible charge storage mechanisms for micropores that are wetted at zero potential. If the electrode surface is charged to a positive potential, an equal negative ionic charge can arise through either (a) adsorption of ex-pore anions into the pores, (b) exchange of ex-pore anions for in-pore cations or (c) the expulsion of cations from the pores. Adapted from Ref. 80 with permission from The Royal Society of Chemistry.

The overall picture that emerges from NMR studies of aqueous electrolytes is that, in contrast to organic electrolytes, spontaneous adsorption is more affected by ion-solvent interactions which can differ appreciably for different ions. These effects are still not well understood and require further study to determine their relative contributions to the $\Delta\delta$ value and the adsorbed ion population. Other factors which also require further study are the contributions of any functional groups on the carbon surface which may be (de)protonated by the aqueous solvent, as well as $OH^-$ and $H_3O^+$ ions that may be present alongside the main electrolyte ions.

*7.3 Characterising Electrosorption Mechanisms in Porous Carbons.*

The traditional picture of the mechanism of electrosorption is that counter-ions are absorbed into the carbon micropores to balance the electronic charge within the carbon surface. However, as discussed above, NMR measurements have clearly shown that there can be a high density of both counter-ions and co-ions (ions with the same charge as the carbon surface) already present inside the pores before a voltage is applied. This means there are a number of possible mechanisms that can lead to the build-up of ionic charge inside the micropores, as illustrated in Fig. 25.[80, 98] Counter-ions can be adsorbed to balance each unit of electronic charge on the electrode surface (Fig. 25a) resulting in an overall increase in the in-pore ion population. However, it is also possible to swap in-pore co-ions for ex-pore counter-ions, as shown in Fig. 25b. In this case, each ion swap balances two units of electronic charge on the electrode surface and the total population of in-pore ions does not change. A further possible mechanism is for co-ions to desorb and be expelled from the pores, leaving a surplus of in-pore counter-ions as shown in Fig. 25c. In this case, one in-pore co-ion is expelled for each unit of charge on the electrode surface and the total in-pore ion population reduced. It is also possible for combinations of these charging mechanisms to take place, or for different charging mechanisms to take place in different electrodes.

In principle, the different possible electrosorption mechanisms can be distinguished and characterised by NMR spectroscopy. This can be done by tracking changes in the in-pore resonance intensity as the electrode is charged, and then converting these into changes in the in-pore ion population. However, this relatively simple concept presents some practical challenges due to the requirement to perform NMR experiments on carbon-electrolyte systems at a defined voltage or charge state. One way that this can be done is to disassemble electrochemical cells and perform *ex situ* MAS measurements on the extracted electrodes. This method has the advantages of providing high resolution spectra and the possibility to perform more sophisticated experiments; however, the electrode voltage cannot be directly measured during the NMR experiment and therefore care must be taken to mitigate against any self-discharge. Another approach is to perform *in situ* NMR measurements using specially designed electrochemical cells located within the NMR detection coil. These experiments are typically carried out under constant voltage conditions, *i.e.*, a fixed voltage is applied to the cell and NMR spectra are acquired after electrochemical equilibriation of the system (a detailed description of the *in situ* NMR methodology and cell designs can be found elsewhere[88, 99, 100]). *In situ* NMR has the advantage that the electrode polarisation is directly known and controlled during the NMR experiment, and a range of electrochemical conditions can be used (*e.g.*, constant voltage, constant current, or cyclic voltammetric conditions). However, as discussed in Section 3.5, static measurements generally show lower resolution than MAS measurements. Also, the cell components can give rise to strong background signals depending on the nucleus studied. The choice of experimental methodology therefore depends upon the properties of the system that is being studied and the information that needs to be obtained.

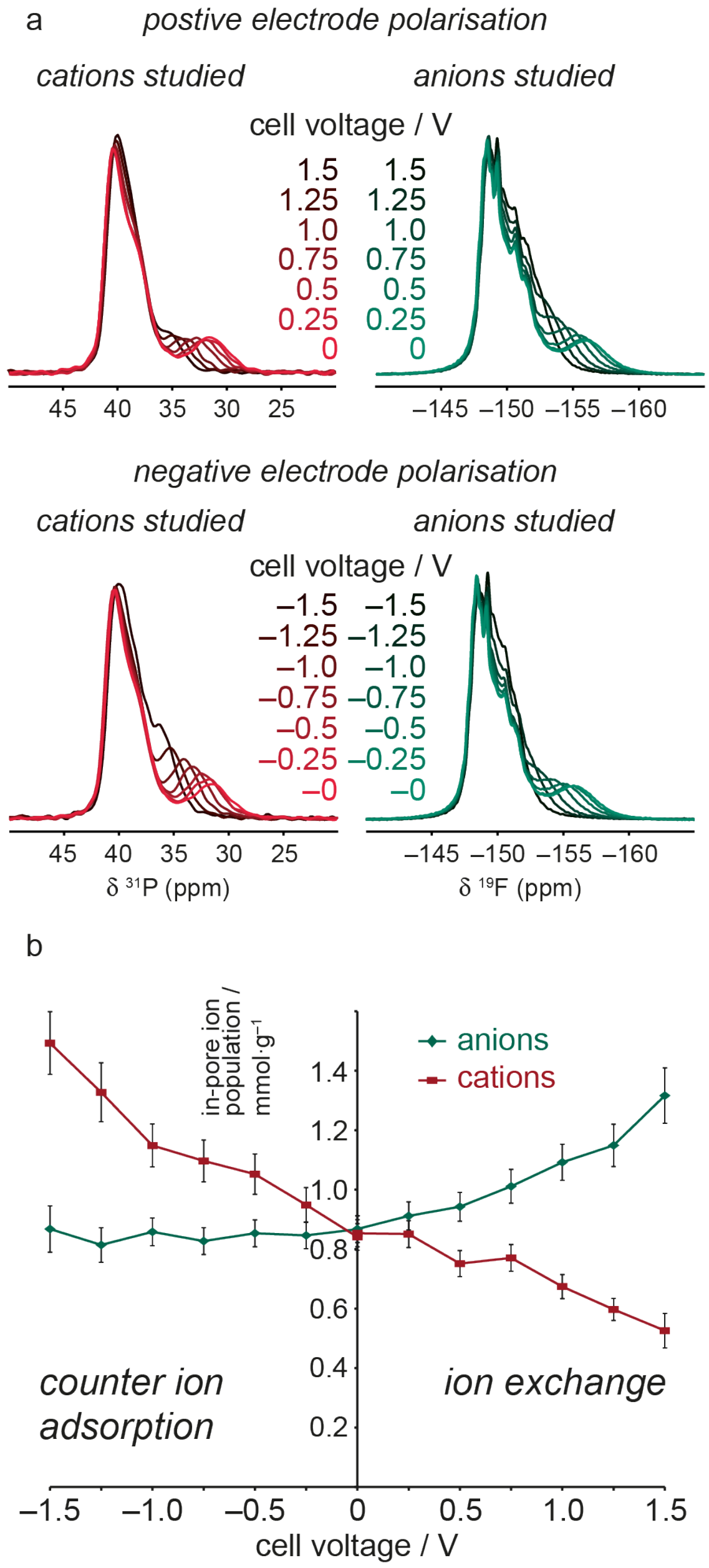

**Figure 26.** (a) $^{31}P$ (cations) and $^{19}F$ (anions) in situ NMR spectra of a 1.5 M $PEt_4$-$BF_4$ / ACN YP50F supercapacitor electrode. Changes in the in-pore resonance intensities are observed as the cell is charged between –1.5 and +1.5 V. (b) In-pore ion populations determined from deconvolutions of the *in situ* NMR spectra reveal a counter-ion adsorption mechanism for negative polarisation and an ion-exchange mechanism for positive polarisation. Reproduced with permission from Ref. 101.

Figure 26a shows *in situ* NMR spectra of a cell containing 1.5 M tetraethylphosphonium ($PEt_4$-$BF_4$) / ACN electrolyte and YP50F activated carbon electrodes.[101] This electrolyte was chosen to facilitate observation of the cations (by $^{31}P$ NMR) as well as the anions (by $^{19}F$ NMR) using nuclei that were not susceptible to background signals from the *in situ* cell components. As the cell is charged, clear changes in the in-pore resonance intensities can be seen, as well as chemical shift changes (discussed later). The in-pore anion intensity is observed to increase for positive voltages, while the in-pore cation intensity increases for negative voltages, consistent with the basic idea of electrosorption of counter-ions. However, when the deconvoluted in-pore resonance intensities are plotted as a function of cell voltage (Figure 26b), greater detail emerges. For positive cell voltages, an ion-exchange mechanism is observed as the in-pore cation population is found to decrease at the same time as the anion population increases. For negative cell voltages, a different mechanism of counter-ion adsorption is observed: the in-pore cation population increases but the in-pore anion population stays approximately constant. Despite the different electrosorption mechanisms, the total in-pore ionic charge (as determined from the sum of in-pore resonance intensities for both ionic species) balances the electronic charge in both electrodes.

The reasons why the electrosorption mechanism for $PEt_4$-$BF_4$ / ACN on YP50F differs with electrode polarity are still not understood. Interestingly, measurements on two lower electrolyte concentrations showed very similar mechanisms, suggesting this is not an influencing factor under the experimental conditions used. *In situ* NMR studies on the very similar $NEt_4$-$BF_4$ / ACN showed a reduction in the in-pore anion population at negative voltages, suggesting a greater contribution of ion exchange for this electrolyte.[99] This was mirrored by *ex situ* NMR measurements by Deschamps *et al.* on both the cations and anions, which revealed ion-exchange processes in both the positive and negative electrodes for two different activated carbons.[102] Other NMR studies have shown that the properties of the electrolyte ions can have an impact on the electrosorption mechanism. *Ex situ* NMR experiments on the ionic liquid $Pyr_{13}$−TFSI adsorbed on YP50F showed that for positive voltages electrosorption took place by ion exchange and counter-ion adsorption, while for negative voltages ion exchange and co-ion desorption were observed.[44] The increased contribution of the TFSI anion to the electrosorption mechanism contrasts $PEt_4$-$BF_4$ / ACN where the anion plays no significant role for negative voltages. In a separate $^{19}F$ *in situ* NMR study, the TFSI anions in Li−TFSI / ACN and Na−TFSI / ACN electrolytes also underwent significant co-ion desorption at negative voltages.[80] These results could suggest that the TFSI anion is more mobile, packs differently, or has different interactions with the carbon surface as compared to $BF_4^-$, although further work is required to investigate this. However, in the same work it was found that desorption of $BF_4^-$ at negative voltages can be promoted by changing the cation to much larger tetrabutylammonium ($NBu_4^+$). It was suggested that the larger cation size restricts their adsorption within small pores, meaning that the smaller $BF_4$ anions are forced to make a greater contribution to the electrosorption mechanism. Similar observations in recent work by Zhang *et al.* support this interpretation and show how it can be exploited to make a supercapacitor device with diode-like properties.[103] Comparison of NMR spectra for 1 M $NBu_4$-$BF_4$ / ACN adsorbed on small pore (0.87 nm) and large pore (4.0 nm) carbons showed that spontaneous adsorption of both $NBu_4$ cations and $BF_4$ anions is prevented in the small pore carbon due to the large cation size. However, the much smaller $BF_4$ anions can be electrosorbed when the small pore carbon is positively polarised. This enabled the design of a "CAPode" device consisting of a large-pore electrode coupled with a small-pore electrode which only stores charge when the small pore electrode is positively polarised and does not function when the polarity is reversed.

Electrosorption of aqueous electrolytes has also been studied by NMR. Luo *et al.* used *in situ* NMR of a cell consisting of 0.8 mol kg$^{-1}$ NaF (aq) and PDC electrodes.[93] For a small-pore PDC (average pore size 0.58 nm), spontaneous adsorption was prevented due to the large solvated ion diameters (as discussed above). As the cell was charged, in-pore resonances in the $^{19}F$ NMR spectrum were not observed until 0.4 V, after which point they increased in intensity consistent with counter-ion adsorption. Following discharge to 0 V, an in-pore resonance remained in the NMR spectrum, suggesting hysteresis in the electrosorption mechanism. These results point towards highly restricted movement of the strongly-solvated electrolyte ions in and out of the small micropores of the carbon electrodes, such that the "gating voltage" of 0.4 V is required to force the ions to partially desolvate and enter the pores. This further highlights the importance of the solvated ion size for the adsorption and electrosorption of aqueous electrolytes. Further work is required to better understand the effect of ion-solvent interactions on the electrosorption mechanism, and how these differ between aqueous and organic electrolytes.

In addition to probing quantities of adsorbed ions at various applied potentials with NMR spectroscopy, it is also possible to realise magnetic resonance imaging (MRI) of working supercapacitors. Ilott *et al.* have reported the first MRI study on a supercapacitor with $NEt_4$-$BF_4$ / ACN as the electrolyte and YP50F as the carbon electrode in 2014.[104] Oukali *et al.* have used *in situ* MRI to compare the electrochemical behaviour of the same electrolyte in two different porous carbons, namely a conventional activated carbon, available commercially, and a CDC.[105] They have shown that the variations in the quantities of adsorbed anions are more dramatic for the CDC, with smaller and more uniform pore sizes, than for the conventional activated carbon, with a number of pores larger than 2 nm, suggesting a stronger interaction between the ions and the carbon in microporous carbons.

*7.4 Chemical Shift Changes due to Electrosorption.*

In addition to changing the in-pore resonance intensity, electrosorption also causes the chemical shifts of in-pore species to change. In general, the in-pore resonance moves to higher chemical shift, *i.e.*, to reduce the magnitude of the $\Delta\delta$ value by several ppm (as shown in Figure 26a). A seemingly obvious explanation for this could be that changing ion-carbon distances during electrosorption cause the $\Delta\delta$ value to change due to the distance dependence of the ring current shielding effect. However, a number of observations suggest that this is not the cause, or at least is not the main contribution to the observed shift changes. Firstly, the direction of the shift change is almost always the same (to higher chemical shift), regardless of the polarity of the electrode. The shift change is also approximately equal for anions and cations. While the precise electrosorption mechanism can be complex (as discussed above), ion-carbon and solvent-carbon distances would not be expected to vary in the same way in electrodes of opposite polarity. Secondly, in an *ex situ* NMR study of the electrosorption of $Pyr_{13}$-TFSI ionic liquid on YP50F, the $^{19}F$ in-pore resonance for the TFSI anion shifted to a higher chemical shift than the ex-pore electrolyte at the highest voltage of 2.5 V.[44] This cannot be explained by changes in ion-carbon distances, which could only reduce the NICS to zero (*i.e.*, to coincide with the ex-pore chemical shift) in the limit of large ion-carbon distance. Furthermore, DFT calculations and lattice simulations have shown that changes of the ion-carbon distance within a micropore cannot account for the magnitudes of the chemical shift changes that are typically observed.[17] Instead, the chemical shift change is thought to be due to the changes in the electronic structure of the carbon surface as the electrode is charged. As explained in Section 3, the exact magnitude and sign of the ring current shift depend precisely on the electronic structure of the carbon. During charging, the unpaired electrons or holes that accumulate in the carbon surface result in an increasingly paramagnetic contribution to the ring current. This cause resonances for species near the carbon surface to move towards higher chemical shift in the NMR spectrum. Importantly, this model predicts that the ring current should become increasingly paramagnetic as more charge is added to the carbon surface, eventually leading to a net paramagnetic (rather than diamagnetic) NICS. This explains the shift of the $^{19}F$ in-pore resonance for the TFSI anion to higher chemical shift than the ex-pore electrolyte in the aforementioned study.[44]

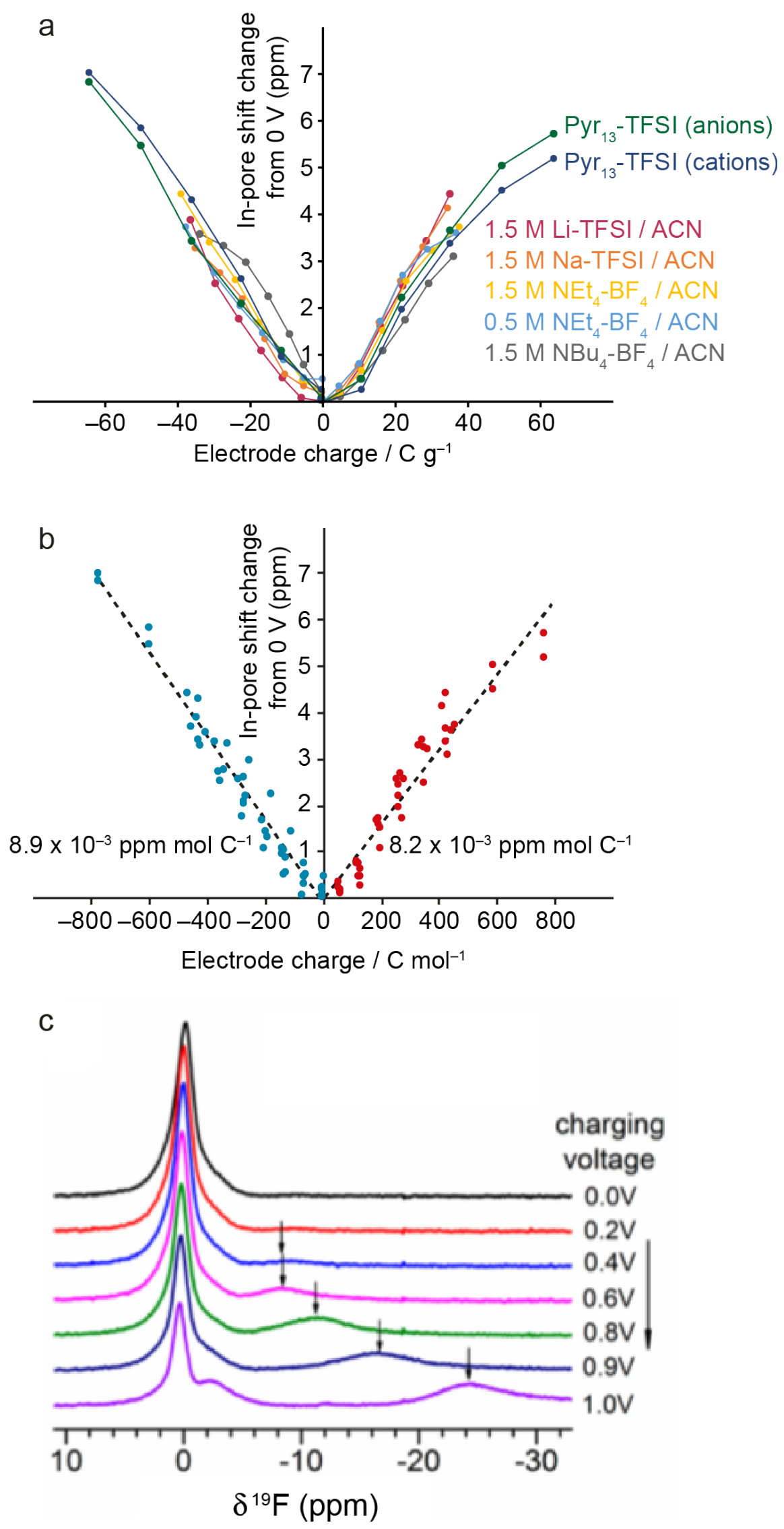

**Figure 27.** (a) In-pore chemical shift changes during electrosorption as a function of applied electrode charge derived from experimental NMR data.[44,80] For ACN-based electrolytes, the anions were studied using $^{19}F$ NMR. For $Pyr_{13}$-TFSI, anions were studied by $^{19}F$ NMR and cations were studied by $^{1}H$ NMR. (b) Linear fits correlating positive and negative charge-induced chemical shift changes with the electrode charge per mole of carbon. (c) Variation of the $^{19}F$ in-pore chemical shift with cell potential in an in situ NMR experiment on NaF (0.8 mol $kg^{-1}$, aq) adsorbed on a PDC with average pore size 0.58 nm [Adapted with permission from Ref. 93. Copyright 2015 American Chemical Society].

The paramagnetic contribution to the NICS for charged carbon surfaces has been qualitatively confirmed by DFT calculations which show that the ring current associated with a small carbon fragment changes sign from diamagnetic to strongly paramagnetic when either a positive or negative charge is applied.[99] For the coronene molecule studied, the NICS calculated at a distance of 4 Å from the carbon surface changed from –3 ppm for a neutral molecule to +5 and +11 for positively and negatively charged molecules, respectively. This qualitatively mirrors experimental observations for [18]-annulene, where the $^{1}H$ chemical shifts of the interior protons changed from –3.0 ppm to 28.1 and 29.5 ppm following two electron reduction.[106]

Figure 27a shows experimental chemical shift changes for a range of electrolytes electrosorbed on YP50 as a function of the integrated electronic charge. For each electrolyte, similar chemical shift changes are observed over the range studied, supporting the idea that the effect is dominated by changes in the electronic structure of the carbon surface, rather than rearrangements or displacements of ions. Despite this, many details of the charge-induced shift change remain poorly understood and further work is required to determine the origin of this phenomenon in more detail, as well as the precise relationship between the ring current-induced shift change and the charge on the carbon surface. In particular, the experimental data in Figure 27a shows the shift change is approximately equal for positive and negative polarisation. This does not agree with the aforementioned DFT calculations which predict that the ring current-induced shift change for a negatively-charged carbon fragment should be approximately two times as large as for a positively-charged fragment.[99] It is not clear if the predicted differences indicate a real effect or a limitation of the DFT method used.

The DFT calculations also appear to underestimate the charge-induced shift change as compared to the experimental data that is available. Figure 27b shows the experimental charge-induced changes in $\Delta\delta$ with the charge density expressed in terms of coulombs per mole of carbon atoms in the electrode. Linear fits to these data give positive and negative charge-induced $\Delta\delta$ variations of +8.2 x $10^{-3}$ and +8.9 x $10^{-3}$ ppm mol $C^{-1}$, respectively. Considering the coronene molecule used for the DFT calculations in Ref. [99], at 4 Å above the carbon surface a charge of $+1e$ or $-1e$ changes the calculated NICS by +8 or +14 ppm, respectively, as compared to the neutral molecule. When expressed per mole of carbon atoms in the coronene molecule, these values correspond to respective charge-induced NICS changes of +2.0 x $10^{-3}$ and +3.5 x $10^{-3}$ ppm mol $C^{-1}$, which are a factor of approximately 3 smaller than the experimental $\Delta\delta$ variation. However, the simple comparison performed here has several limitations which may contribute to this discrepancy. Although expressing the charge density in coulombs per mole of carbon atoms allows for a straightforward comparison between the experimental data and the model coronene molecule, it does not capture possible heterogeneity in the electronic charge distribution which is likely to be concentrated at pore surfaces. In this respect, normalising the charge density by surface area may give a better comparison, although surface areas of activated carbons are difficult to accurately define. Another limitation is that the DFT NICS calculations were performed above a single charged carbon surface. As discussed in Section 4, better agreement has been obtained for uncharged systems when a model slit pore comprising two carbon surfaces is used, particularly when the effects of intra-pore exchange dynamics are included.[17] In this case, the NICS from both surfaces are additive, so larger shift changes would be predicted. Additionally, the charged DFT calculations were performed on coronene whereas calculated NICS values for neutral circumcoronene have been found to give better agreement with $\Delta\delta$ for several carbons.[48] Curvature of the carbon surface may also influence the magnitude of the charge-induced

NICS change. Further work is required to investigate to what extent each of these factors contribute charge-induced shift changes, as well as to determine any other effects that may be important.

While the charge-induced shift change is almost always to higher chemical shift, Luo *et al.*, have observed an exception to this trend in the electrosorption of NaF (0.8 mol $kg^{-1}$ aq) electrolyte on a small-pore PDC.[93] In an *in situ* NMR study of this system, a "gating voltage" of 0.4 V was required to observe entry of $F^-$ ions (0.7 nm diameter) into the 0.58 nm pores, as discussed above. Between 0.4 – 0.7 V the magnitude of the $\Delta\delta$ value was approximately constant at 9 ppm, but as the voltage was increased, the $\Delta\delta$ value increased, reaching –25 ppm at the maximum voltage of 1.0 V (Figure 27c). This chemical shift change is in the opposite (diamagnetic) direction to what is typically observed, and is also much larger, suggesting that it comes from a different mechanism. This was attributed to partial dehydration of the $F^-$ ions upon pore entry, which is necessary due to the small pore width in the carbon used. The authors cited a DFT study which has shown that the removal of up to two water molecules from the hydration shell can lead to a 13 ppm chemical shift change, which is consistent with the observed shift change between 0.7 – 1.0 V.[97] In the same study, the electrosorption of cations on the same PDC was also tracked by $^{23}$Na NMR. Interestingly, while an even higher gating voltage of 0.6 V was required for entry of the Na cations into the pores, no large shift change was observed, suggesting that the hydration shell is not significantly disrupted upon pore entry. Clearly, more work must be done to disentangle the different factors that influence charge-dependent chemical shifts in porous carbon materials.

## 8. Summary and Outlook

Over the last 25 years, NMR spectroscopy has emerged as a powerful tool for probing adsorption and diffusion in carbonaceous materials. The primary reason for this is the unique ability to resolve adsorbed species from those outside the carbon pores due to the ring current shift. However, since the initial observations of this phenomenon, many advanced experimental and methodological approaches have been developed that go far beyond basic one-dimensional spectra, enabling more detailed information to be gained. To some extent, the development of NMR methods for studying porous carbon systems in recent years has been driven by their increasing importance in a variety of practical applications such as energy storage and water purification. However, much of the knowledge that has been gained through the study of porous carbons by NMR has contributed to the understanding of fundamental phenomena such as ion confinement, ion transport and carbon aromaticity.

While the range of NMR approaches that have been developed have provided lots of important information, there are many future directions in which further development and understanding is required (Table 4). One such direction is to further study the factors affecting the magnitude of the $\Delta\delta$ value for adsorbed species. Although the distance dependence from the carbon surface is possibly the most obvious factor, in recent years there has been a growing awareness of other factors such as exchange effects (and the related influence of particle size), desolvation, host-guest interactions and aromaticity within the carbon structure. Future systematic studies of these effects will help to better understand their relative contributions as well as how they may be used to provide structural information about porous carbons and/or structural and dynamic information about adsorbed species. In addition, the marked changes in the $\Delta\delta$ values that have been observed for charged carbon surfaces remain poorly understood. Although a number of systems have now been studied under applied potentials, further experimental studies will help to build up a picture of the generality of the observed transition from diamagnetic to paramagnetic ring current shift. In particular, theoretical modelling will also play an important role in rationalising the observed changes in terms of the electronic structure of the carbon surface.

A further area for future investigation is the differences in adsorption behaviour between organic electrolytes, ionic liquids and aqueous electrolytes. Owing to the recent interest in supercapacitor energy storage devices, much of the work done so far has focused on organic electrolytes. The minority of studies on aqueous electrolytes suggest that the adsorption and electrosorption behaviour is markedly different, possibly due to different ion-solvent interactions. Further experimental and theoretical work will help to understand and quantify these differences, both on spontaneous adsorption and under applied potentials.

**Table 4.** Selected key findings presented in this review, and some remaining related challenges.

| Key Findings | Remaining Challenges |
|---|---|
| Molecules adsorbed in porous carbon materials show a distinct "in-pore" resonance, with a chemical shift that is generally negative relative to the neat adsorbate (negative $\Delta\delta$ value). The ring current effect (which can be partly estimated as a NICS) is a key contribution to the $\Delta\delta$ value. | Understanding the full range of factors that determine the $\Delta\delta$ values beyond the NICS term.<br><br>Resolution of the in-pore resonance can be poor in the NMR spectra of some samples. |
| For a given carbon adsorbent, similar $\Delta\delta$ values are observed for different adsorbates (Table 1). | Understanding what gives rise to the (generally) small differences in $\Delta\delta$ values for different adsorbates (Table 1). |
| For different carbons, different $\Delta\delta$ values are observed for a given adsorbate. Structural information on carbon pore size and carbon ordering / local structure is accessible. | Properly establishing the link between heterogeneous carbon structures and the NMR spectrum for adsorbed species. |
| Chemical exchange between in-pore and ex-pore molecules can have a significant impact on the spectral lineshapes and chemical shifts, and these exchange processes can be quantified using exchange spectroscopy experiments. Interpore (i.e., pore to pore) exchange can also impact the lineshape. | Quantifying the amount of adsorbed species (in-pore species) in systems with complex exchange processes, and quantifying and understanding the different dynamic processes. |
| Pulsed field gradient NMR experiments enable access to in-pore self-diffusion coefficient, which are greatly reduced compared to self-diffusion in the bulk fluids. In-pore diffusion generally becomes slower for carbons with smaller pore sizes. | Understanding the link between carbon structure/porosity and the in-pore diffusion.<br><br>Understanding the role of hierarchy in the pore network on diffusion. |
| Upon charging a carbon material, the electrosorption of ions can be quantified using NMR spectroscopy. Different charging mechanisms can be distinguished involved ion-exchange, ion adsorption, or ion desorption/ejection. | Accurately quantifying the in-pore resonance in charging experiments that often yield spectra with poor resolution.<br><br>Understanding the link between carbon and electrolyte structure on charging mechanisms. |
| Charging a carbon material brings about changes to the $\Delta\delta$ value for in-pore species. The changes in $\Delta\delta$ value appear to be dominated by changes to the carbon electronic structure, in turn altering the ring currents. | Understanding the link between the carbon electronic structure, and the observed chemical shifts.<br><br>Understanding other factors, such as solvation, that may influence the charge dependence of $\Delta\delta$ values. |